\documentclass[11pt]{article}

\usepackage{amsmath,amsfonts}
\usepackage{subfigure}
\usepackage{amssymb,amsthm,setspace}
\usepackage{graphicx}

\usepackage{xcolor}
\newcommand\crule[3][black]{\textcolor{#1}{\rule{#2}{#3}}}

\newcommand{\be}{\begin{equation}}
\newcommand{\ee}{\end{equation}}

\begin{document}

\title{Wave propagation across interfaces induced by different interaction exponents in ordered and disordered Hertz-like granular chains}

\author{Armin Keki\'c and Robert A. Van Gorder \\  \small  Mathematical Institute, University of Oxford\\ \small Andrew Wiles Building, Radcliffe Observatory Quarter, Woodstock Road \\ \small Oxford, OX2 6GG, United Kingdom\\
\small Email: Robert.VanGorder@maths.ox.ac.uk}        
\date{\today}

\maketitle

\begin{center}
Abstract
\end{center}
We study solitary wave propagation in 1D granular crystals with Hertz-like interaction potentials. We consider interfaces between media with different exponents in the interaction potential. For an interface with increasing interaction potential exponent along the propagation direction we obtain mainly transmission with delayed secondary transmitted and reflected pulses. For interfaces with decreasing interaction potential exponent we observe both significant reflection and transmission of the solitary wave, where the transmitted part of the wave forms a multipulse structure. We also investigate impurities consisting of beads with different interaction exponents compared to the media they are embedded in, and we find that the impurities cause both reflection and transmission, including the formation of multipulse structures, independent of whether the exponent in the impurities is smaller than in the surrounding media. We explain wave propagation effects at interfaces and impurities in terms of quasi-particle collisions. Next we consider wave propagation along Hertz-like granular chains of beads in the presence of disorder and periodicity in the interaction exponents present in the Hertz-like potential, modelling, for instance, inhomogeneity in the contact geometry between beads in the granular chain. We find that solitary waves in media with randomised interaction exponents (which models disorder in the contact geometry) experience exponential decay, where the dependence of the decay rate is similar to the case of randomised bead masses. In the periodic case of chains with interaction exponents alternating between two fixed values, we find qualitatively different propagation properties depending on the choice of the two exponents. In particular, we find regimes with either exponential decay or stable solitary wave propagation with pairwise collective behaviour. For some specific interaction exponent values, we observe a new type of stable confined wave which exhibits a periodically changing wave form. These results cast light on how inhomogeneity in the contact geometry will influence solitary wave propagation along Hertz-like granular chains.
\\
\\
\noindent \textit{Keywords}: granular chain; Hertzian contact law; highly nonlinear model; interfaces; interaction exponents; solitary waves

\section{Introduction and mathematical model}
Granular materials are composed of a large number of discrete, solid macroscopic particles such as sand, rice or snow; and they play an important role in many aspects of industry and science such as construction, agriculture or geological processes. They are often viewed as a fourth state of matter as they can exhibit characteristics of gases, liquids or solids, depending on the circumstances \cite{JAE01}. One of the most interesting features of granular crystals we want to study is wave propagation. Compression waves can propagate as solitary or sound waves depending on the interaction potential between the beads and the external compression of the chain \cite{SEN01,chong2017nonlinear,rosas2018pulse}. In uniform granular crystals solitary waves can travel over distances that are orders of magnitude larger than the wavelength without experiencing any change in shape due to dispersion \cite{NES02}. One type of interaction that leads to non-linear solitary waves is the repulsive Hertz-like contact force that depends on the contact geometry and bead material \cite{JOH01}. Even though solitary waves are stable in uniform granular chains \cite{NES02}, they can undergo drastic changes at interfaces between media with different properties \cite{SEN01} or at boundaries \cite{job2005hertzian}. The wave can be partially transmitted or reflected, and even the full disintegration of the wave into so-called multipulse structures is observed \cite{VER01, VER02}. While most of the work done on interfaces so far focuses on media with different bead masses, we use differences in the contact geometry between the beads as an alternative approach to tweak the wave propagation dynamics at interfaces. Indeed, the shape of the interaction potential depends on the geometry of the contact region, which in turn affects the wave propagation \cite{SEN01}. 

\begin{figure}[h]
\begin{center}
\includegraphics[width=0.45\textwidth]{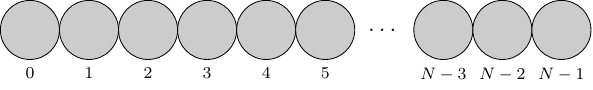}
\vspace{-0.1in}
\caption{Granular chain composed of uniform media. Note that in the experimental implementation the beads are not spherical in general.}
\label{fig:chain}
\end{center}
\end{figure}

The interesting wave propagation properties in one-dimensional a granular chain, as depicted in Figure \ref{fig:chain}, originate in the interaction between neighbouring beads in the chain. Two elastic bodies in contact experience a repulsive force which depends on the compression. Considering two bodies $i$ and $i+1$ at positions $x_i$ and $x_{i+1}$, the repulsive force can be obtained by the interaction potential
\begin{equation}
  V(\delta_{i, i+1}) = a_{i,i+1} \delta_{i,i+1}^{n_{i,i+1}} ,
  \label{eqn:potential}
\end{equation}
where $\delta_{i,i+1}$ is the virtual overlap between the two bodies presuming they would not deform, and $a_{i,i+1}$ is a prefactor depending on the elastic properties of the material and the geometry of the contact region \cite{JOH01, KHA01}. We assume that the mutual force between beads is given only by contact forces and therefore $V\equiv0$ for $\delta_{i,i+1}=0$, i.e. there is no long-range interaction. The value of the exponent $n_{i,i+1}$ is determined by the contact region geometry \cite{JOH01, SEN01}.
\par
The repulsive force between beads is related to the interaction potential by
\begin{equation}
   F(\delta) = -\frac{\partial}{\partial \delta} V(\delta).
\end{equation}
We only consider interactions with $n_{i,i+1}>2$, for which the forces between the beads are non-linear in $\delta$. This leads to the formation of non-linear waves, for which the propagation speed depends on the amplitude. Furthermore, we note that the inter-particle force originating from the potential \eqref{eqn:potential} is fully non-linear, i.e. the force has no linear component. As a direct consequence of that, sound wave propagation is not possible in granular crystals without external precompression \cite{NES02}. Such media are sometimes referred to as \textit{sonic vacua} \cite{NES02}.

For spheres with radii $R_{i}$ and $R_{i+1}$ the exponent is $n_{i,i+1}=5/2$ and the prefactor is given by
\begin{equation}
  V(\delta_{i, i+1}) \stackrel{\text{\tiny{spheres}}}{=} \frac{2}{5 D_{i,i+1}} \sqrt{\frac{R_i R_{i+1}}{R_i+R_{i+1}}} \delta_{i,i+1}^{5/2} ,
  \label{eqn:potential2}
\end{equation}
\begin{equation}
	D_{i,i+1} = \frac{3}{4} \left[ \frac{1-\sigma_i^2}{Y_i} + \frac{1-\sigma_{i+1}^2}{Y_{i+1}} \right] ,
	\label{eqn:prefactor}
\end{equation}
where the Young moduli $Y_i$, $Y_{i+1}$ and the Poisson ratios $\sigma_i$, $\sigma_{i+1}$ are determined by the elastic properties of the bead materials. Equation \eqref{eqn:potential2} is called \textit{Hertz potential} \cite{HER01, LAN01}; systems with the more general interaction potential \eqref{eqn:potential} are sometimes called \textit{Hertz-like}. Note that chains of hollow spheres can also be modeled in the $n = 5/2$ regime \cite{vorotnikov2017wave}. 

The overlap for spherical beads is given by
\begin{equation}
  \delta_{i,i+1} = \left( \left(R_i + R_{i+1} \right) - x_{i+1} + x_i \right)_+ ,
\end{equation}
with $v_+ := \max(v,0)$. It is often more convenient to write the overlap in terms of the displacements rather than the absolute positions of the beads. For bead $i$ the displacement is $u_i = x_i - x_{i,0}$, where $x_{i,0}$ is the initial position, and the overlap becomes 
\begin{equation}
  \delta_{i,i+1} = \left( \Delta_{i,i+1} - u_{i+1} + u_i \right)_+ .
\end{equation}
The precompression $\Delta_{i,i+1}$ accounts for the fact that the initial position of the beads need not be such that the beads are barely touching, i.e. touching but not overlapping. They can also be precompressed by squeezing the chain from both ends.

The potential \eqref{eqn:potential} is, strictly speaking, a result for the static case, i.e. Equation \eqref{eqn:potential} relates the overlap between the beads to an externally applied static force. This result only remains valid for the dynamic case if the time scale of the change in force is much larger than the time that a longitudinal acoustic wave would need to travel across the length of the bead \cite{COS01}. This restriction ensures that the internal dynamics of the bead can be neglected for the study of the dynamics in the chain. We use similar chain parameters as in the experimental study carried out by Coste et al.\ \cite{COS01} in order ensure the validity of this approximation. Therefore, we can treat the beads as point-like masses that interact via \eqref{eqn:potential}.

A noteworthy property of the type of granular chains we want to study is that they can be torn apart without any resistance, which is a consequence of the fact that there is no long-range interaction between the beads. This is sometimes referred to as the chain having zero \textit{tensile strength}. Therefore, gaps can open between beads in the chain; and without external compression these can only be closed by interacting with other beads.

From Equation \eqref{eqn:potential} one can derive Newton's equation of motion for the $i^\mathrm{th}$ bead, with $i\in\{ 0, \dots, N-1\}$, in a granular chain
\begin{equation}\begin{aligned}
   \frac{d^2 u_i(t)}{d t^2} & = \frac{1}{m_i} \underbrace{W_i}_{\text{external field}}  + \frac{n_{i-1,i} a_{i-1,i}}{m_i} \underbrace{\left( \Delta_{i-1,i} + u_{i-1} (t) -u_i(t) \right)^{n_{i-1,i}-1}_+}_{\text{left neighbour}}  \\
   & \quad - \frac{n_{i,i+1} a_{i,i+1}}{m_i} \underbrace{\left( \Delta_{i,i+1} + u_{i} (t) -u_{i+1}(t) \right)^{n_{i,i+1}-1}_+}_{\text{right neighbour}}  .
  \label{eqn:eom}
\end{aligned}\end{equation}
Here, the last two terms describe the interaction with the left and right neighbour respectively. In addition to the mutual repulsive contact force described by Equation \eqref{eqn:potential}, an external force field, such as gravity, can be applied to the beads, which is accounted for in the first term. For the beads at the end of the chain the relevant neighbour interaction term needs to be set to zero. Chains in which all the beads and precompressions are equal are called \textit{monodisperse}, and \textit{polydisperse} otherwise. We call a bead \textit{excited} if it has significant overlap with at least one of its neighbours; in other words, if there is some potential energy stored in the interaction with a neighbouring bead.

\begin{figure}
\begin{center}
\includegraphics[width=0.35\textwidth]{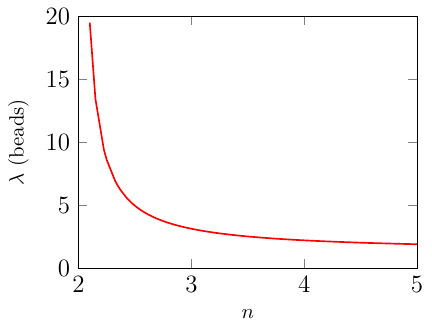}
\vspace{-0.2in}
\caption{Wavelength $\lambda$ as function of $n$ as predicted by continuum model.}
\label{fig:wavelength}
\end{center}
\end{figure}

The existence of solitary waves in granular chains was first theoretically predicted using a continuum model by Nesterenko \cite{NES03} and experimentally verified by Lazardi and Nesternko \cite{LAZ01} and later by Coste et al. \cite{COS01}. The analytical formulation for the propagation of a single solitary wave through a granular crystal under the assumption of a continuous medium \cite{NES02, SEN01} shows good agreement with experiments and numerical simulations for simple solitary wave propagation in monodisperse media \cite{HAS01, LAZ01, COS01}. Other analytical approximations have been obtained by \cite{chatterjee1999asymptotic,hasan2017basic,tang2017novel}. The wavelength as a function of $n$ given by the solution of the continuum model \cite{NES03} is shown in Figure \ref{fig:wavelength}. As shown by Hasco\"et et al. \cite{HAS01}, the wavelength in the simulation of the discrete system is in good agreement with the continuum model for large wavelengths, i.e.\ small $n$, and is slightly underestimated for short wavelengths, i.e.\ large $n$. Note that in the limit of large $n$ only two beads interact at a given point in time forming the solitary wave.

However, the continuum model fails to predict the wave propagation behaviour in more complex situations where the discreteness of granular media can play an essential role. For example, two colliding solitary waves pass each other without changing their shape and amplitude in continuous media, whereas numerical simulations of the discrete medium have shown the emergence of secondary solitary waves from the collision \cite{SEN01, MAN01, FEL01}.

The simplest form of a polydisperse chain is one which consists of two jointed monodisperse granular chains with different bead masses. The behaviour of a solitary wave propagating across the interface between the two media depends on the side from which the wave initially approaches the interface. In numerical investigations, the partial transmission and reflection of the incoming wave are observed for a solitary wave propagating across an interface from a medium with light grains to one with heavier grains \cite{VER01, VER02, SEN01}. Furthermore, during the splitting of the incoming wave into a transmitted and a reflected part, gaps arise between beads at the vicinity of the interface \cite{VER02}. Similar effects to those at interfaces between granular crystals with different bead masses have been observed for mass impurities. These impurities are created by changing the mass of a single bead in an otherwise monodisperse granular crystal \cite{HAS01, SEN01, HON02}. For a wave travelling from a medium with heavy grains to one with lighter grains in a jointed chain, the incoming wave is fully transmitted across the interface and disintegrates forming a multipulse structure \cite{NES01, VER02}. A second multipulse structure is formed with a delay to the first one. This delay is due to a gap opening and closing between beads at the interface. 

Wave propagation across mass interfaces is another example in which effects arise as a direct consequence of the discreteness of the medium, as gaps between beads in the vicinity of the interface are crucial for understanding the wave dynamics \cite{VER02, NES02}. These phenomena cannot be captured by the analytical approaches used to describe granular crystals developed so far, due to the assumption of a continuous medium used in such analytical approaches. Consequently, the study of wave propagation across interfaces is limited to numerical simulations.

Granular media have received a lot of attention from material scientists due to their application in the design of impact attenuating materials \cite{POR01,BUR01, GAN01, CHA01, LEO01, MUE01}. In order to properly make use of the shock absorbing capabilities of granular crystals, understanding the propagation of solitary waves is indispensable. In granular crystals, designing an impact attenuating material is equivalent to designing a medium that disintegrates or re-directs solitary waves in such a way that the impact of the wave on the area which is to be protected is mitigated. In 1D, so-called \textit{granular containers} make use of the propagation of waves across mass interfaces in the design of a composite material that is capable of delaying and mitigating the impact of an incoming solitary wave. The simplest form of a granular container is a medium with lighter beads embedded in a medium with heavier beads \cite{HON01, SEN01, VER02}. At the first interface a solitary wave disintegrates and forms a multipulse structure inside the container, whereas at the second interface each of these pulses is partially transmitted and reflected. This leads to only a fraction of the initial pulse energy being transmitted after the pulse meets the two interfaces. The remaining energy is held back in the container in the form of several solitary waves which are reflected back forth, losing part of their energy in transmission at each reflection. Thus the impact of the initial solitary wave is attenuated by splitting it into many smaller waves which hit the end of the chain one after the other \cite{VER02}. This principle of combining interfaces to manipulate the wave propagation can be further extended to more complex granular container structures \cite{SEN01}.

\textit{Tapered chains} are chains with slowly increasing bead masses; i.e. instead of having one interface with a relatively high difference in bead masses, tapered chains can be seen as a sequence of interfaces. This set-up reflects a fraction of the wave energy at each interface. Tapered chains are characterised by the tapering parameter $q\in (0,1)$ that relates the radii $R_i$ and $R_{i+1}$ of two successive spherical beads in the tapered part of the chain by $R_{i+1} = (1-q)R_i$. In numerical simulations it was observed that the wave attenuation is increased with increasing tapering $q$. The attenuation can be further enhanced in so-called \textit{decorated tapered chains}, where every other bead in the decorated part of the chain is kept at the initial bead size \cite{DON01,SEN01}.

The propagation of solitary waves in granular chains across interfaces between media with different bead masses has been studied extensively \cite{SEN01}. Effects including reflection, transmission and emerging multipulse structures have been found and used to design composite materials for manipulating the propagation behaviour of incoming solitary waves \cite{VER01, VER02, MAN02, NES01, DAR01}.

\begin{figure}[h]
\begin{center}
\includegraphics[width=0.45\textwidth]{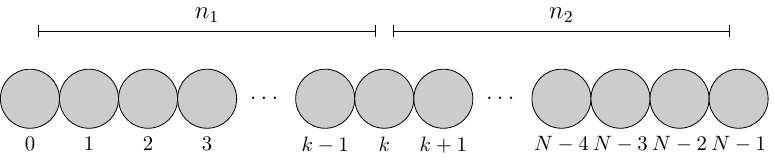}
\vspace{-0.1in}
\caption{Granular chain composed of two media with interaction potential exponents $n_1$ and $n_2$ with the interface being at node $k$. The exponent difference is defined as $\Delta n = n_2 - n_1$. Note that in the experimental implementation the beads are not spherical in general.}
\label{fig:interface}\end{center}
\end{figure}

In this paper we present a new way of creating interfaces between two types of granular crystals that offers an alternative method of affecting solitary wave propagation in composite media. Knowing that differences in bead masses can induce complex dynamics at interfaces, we want to investigate how abrupt changes in the interaction potential influence the wave propagation. While there are some results for interfaces between linear and non-linear media \cite{SEN01}, we are not aware of any work on interfaces between media with different nonlinearities. Therefore, we study the propagation properties for solitary waves travelling across interfaces between media with different interaction exponents $n$. For a schematic, see Figure \ref{fig:interface}. 

The underlying physical property causing the interface effects for mass interfaces is the inertia of the beads, i.e. the resistance to acceleration, for which the bead mass is a measure. These differences in acceleration behaviour lead to a disruption of the compression wave propagation causing the opening and closing of gaps and compression imbalances in the vicinity of the interface. The parameter $n$ governs the shape of the interaction potential and thereby the shape of the interparticle force. And since according to Newton's second law of motion the force is proportional to the acceleration of the particles we expect imbalances in the acceleration behaviour of the beads at the interface. Therefore, disruptions of the solitary wave propagation are expected in this case as well.
Similarly to mass interfaces, the interface effects for different interaction exponents depend on whether the exponent increases or decreases along the propagation direction. We characterise interfaces by the difference $\Delta n = n_2-n_1$ between the exponents in the two media.

We shall also be interested in considering periodicity or disorder in the contact geometry between the beads within a granular chain. In numerical simulations of with slowly increasing bead masses, so-called \textit{tapered chains}, it was observed that the wave attenuation is increased with increasing tapering \cite{DON01,SEN01}. A similar idea to tapering chains is to introduce disorder in the bead masses \cite{SEN01, MAN02}. This acts as a sequence of interfaces; however, here the mass difference along the wave propagation can be both positive or negative. Other systems using a sequence of interfaces are granular crystals with alternating bead masses, so-called \textit{dimer} or \textit{di-atomic} granular crystals \cite{POR02}. Although compression of monodisperse granular chains causes wave dispersion, stable and spatially localised waves was be found for di-atomic chains \cite{BOE01, THE02}. The propagation of solitary waves in granular chains across interfaces between media with different bead masses has been studied extensively \cite{SEN01}, and these studied motivated the design of composite materials for manipulating the propagation behavior of incoming solitary waves \cite{VER01, VER02, MAN02, NES01, DAR01}.

We shall consider the effects on the wave propagation dynamics of changing the interaction exponents on all pairs of beads independently. We introduce these variations in the interaction exponent in two different ways: by randomly selecting exponents, or by periodically alternating between two fixed values. Such configurations can be seen as the limit of many interfaces, where the media between the interfaces are made up of only single beads. 

The remainder of this paper is organized as follows. In Section 2, we outline the mathematical model for a Hertz-like granular chain, and non-dimensionalize it. We also discuss the numerical simulation approach. In Section 3 we consider the case of interfaces formed when granular chains are composed of beads with different interaction exponents in the Hertzian contact law. In Section 4 we consider the scenario where the interaction exponents differ periodically between two fixed values, resulting in a type of periodic media (akin to what has previously been done for bead masses). In Section 5 we consider the case of either local or global disorder in the interaction exponents, randomly selecting the value of the interaction exponents, again akin to what has previously been done for bead masses in \cite{MAN02}. Finally, in Section 6, we comment on wave propagation across interfaces between granular chains composed of beads with both different interaction exponents and different masses. Finally, we offer summarizing remarks and conclusions. 

\section{Non-dimensional model and simulation parameters}
In order to induce waves in the granular chain the initial conditions of the beads are set as 
\begin{equation}
  u_1, \dots, u_N = 0, \quad
  v_1 = v^{(0)}, \quad \text{and} \quad v_2, \dots v_N = 0 ,
  \label{eqn:init}
\end{equation}
where $v_k = \frac{du_k}{dt}$. This corresponds to the first bead hitting the granular chain from the left and thereby exciting beads in the system, resulting in wave propagation.

The beads are assumed to be made of stainless steel and dimensions are as used in the experimental set-up by Coste et al. in \cite{COS01}. Furthermore, it is assumed that the distance between the two contact points of the beads is equal to the bead diameter used in \cite{COS01}. The relevant physical parameters are shown in Table \ref{tab:properties}. 

\begin{table}[htb]
	\begin{center}
	\begin{tabular}{@{}l l l@{}}
	\hline
	Symbol & Property & Value\\
	\hline
	$R$ & bead radius &  $4\text{mm}$ \\
	$Y$ & Young's modulus &  $2.26\times 10^{11}\text{N}\text{m}^{-2}$ \\
	$\sigma$ & Poisson's ratio &  $0.3$ \\
	$\rho$ & density &  $7650 \text{kg}{m}^{-3}$ \\
	$m$ & mass &  ${2.05}\text{g}$ \\
	\hline
	\end{tabular}
	\caption{Physical parameters used in simulations \cite{COS01}.}
	\label{tab:properties}\end{center}
\end{table}

The initial conditions \eqref{eqn:init} are chosen such that the wave speed of the induced solitary wave is in the range of wave speeds experimentally tested in \cite{COS01}. In computer units, i.e. after re-scaling, for spherical beads with masses as in Table \ref{tab:properties} we set the initial velocity of the left striking bead to be  $v^{(0)}=1\times 10^7$. 

The value the initial velocity takes in computer units changes for  different re-scaling factors; however in SI units the initial velocities are the same: approximately $0.11\text{m}\text{s}^{-2}$. This initial velocity induces waves with wave speeds that are of the order of waves experimentally measured by Coste et al. \cite{COS01}.

The interaction potential exponent $n$ can be changed experimentally by altering the contact geometry between the beads. However, changing the contact geometry will generally affect the prefactors $a$ as well \cite{JOH01, LAN01}. For the sake of simplicity, it is assumed that the prefactors are constant over the entire range of exponents tested here. We set the value of the prefactors to the value for spherical beads as in Equations \eqref{eqn:potential2} and \eqref{eqn:prefactor}.

\begin{figure*}
\begin{center}
    \includegraphics[width=0.35\textwidth]{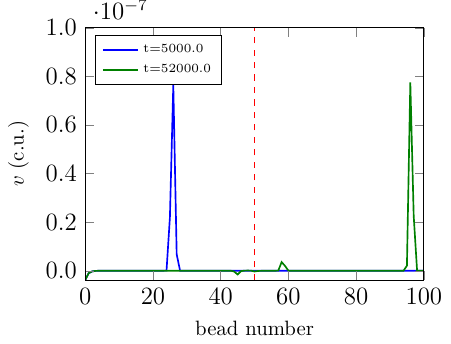}
    \includegraphics[width=0.32\textwidth]{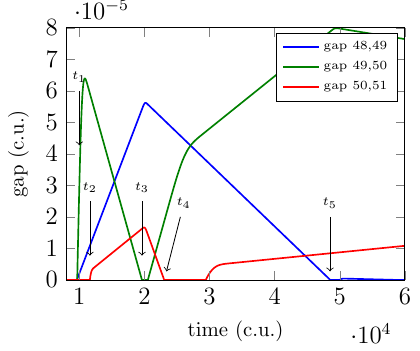}
    \vspace{-0.2in}
       \caption{Solitary wave propagation across an interface between media with $n_1=3.0$ and $n_2=3.5$: waves before and after hitting the interface (top) and gaps emerging in the vicinity of the interface (bottom). In the top panel, the interface is indicated by the dashed red line.}\label{fig3}\end{center}
\end{figure*}
\begin{figure*}
\begin{center}
\includegraphics[width=0.19\textwidth]{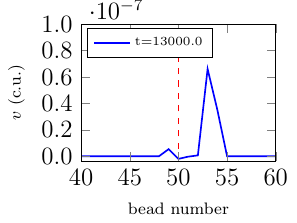}
\includegraphics[width=0.19\textwidth]{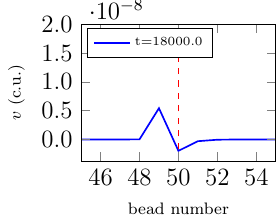}
\includegraphics[width=0.19\textwidth]{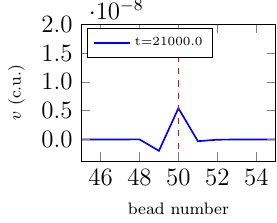}
\includegraphics[width=0.19\textwidth]{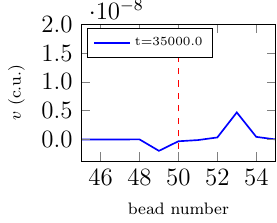}
\includegraphics[width=0.19\textwidth]{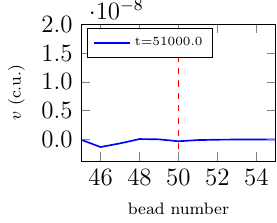}
\vspace{-0.2in}
  \caption{Solitary wave propagation across an interface between media with $n_1=3.0$ and $n_2=3.5$ for different times. The interface is indicated by the dashed red line.}\label{fig4}\end{center}
\end{figure*}%

For typical experimental parameters of the chain \cite{COS01}, the factors scaling like $na/m$ in front of the interaction term in the equation of motion are $\mathcal{O}(10^{12}\text{s}^{-2}\text{m}^{-(n-2)})$ in SI units, since
\begin{equation}
  a=\mathcal{O}\left(10^{9}{\frac{\text{kg}}{\text{s}^{2} \text{m}^{(n-2)} }}\right), \quad m=\mathcal{O}(10^{-3}\text{kg}), \quad n = \mathcal{O}(1).
\end{equation}
In order to scale this factor to $\mathcal{O}(1)$ we re-scale time by 
\begin{equation}
  t = \beta \tilde{t}, \qquad \beta=\sqrt{\frac{m_0}{\mu n_0}} ,
  \label{eqn:scaling}
\end{equation}
where $\mu$ is the order of magnitude of $a$ such that $\tilde{a}=a/\mu=\mathcal{O}(1\text{kg}\text{s}^{-2}\text{m}^{-(n-2)})$, and $m_0$ and $n_0$ are typical values for the bead mass and interaction potential exponent respectively. Then the equation of motion after rescaling becomes
\begin{equation}\begin{aligned}
   \frac{d^2 u_i(\tilde{t})}{d \tilde{t}^2} & = \frac{\beta^2}{m_i} W_i + \frac{\beta^2 \mu n_{i-1,i}}{m_i} \tilde{a}_{i-1,i} \left( \Delta_{i-1,i} + u_{i-1} (\tilde{t}) -u_i(\tilde{t}) \right)^{n_{i-1,i}-1}_+  \\
   & \quad - \frac{\beta^2 \mu n_{i,i+1}}{m_i} \tilde{a}_{i,i+1} \left( \Delta_{i,i+1} + u_{i} (\tilde{t}) -u_{i+1}(\tilde{t}) \right)^{n_{i,i+1}-1}_+  .
  \label{eqn:eom2}
\end{aligned}\end{equation}
Details of the numerical approach and error analysis are provided in Appendix A. Furthermore, the stability of the wave propagation under small perturbations to the interaction exponents (i.e. structural perturbations to the system) is discussed in Appendix B.

\section{Interfaces created by different interaction exponents}

\subsection{Increasing interaction exponent}

Figure \ref{fig3} shows the bead velocities before and after the solitary wave hits an interface with $n_1=3.0$ and $n_2=3.5$. For interfaces with $\Delta n>0$ we observe that most of the energy is transmitted to the medium on the right-hand side; however, there is one additional transmitted pulse and one reflected pulse, both about an order of magnitude smaller in peak velocity. A noteworthy behaviour is that the bead velocities in the vicinity of the interface freeze in a few configurations before all the energy is propagated away from the interface.

Whereas the arriving wave in the medium on the left immediately induces a wave of similar amplitude in the medium on the right, the two secondary pulses are delayed, i.e. they do not propagate away from the interface directly after the initial solitary wave hits it, but they freeze close to the interface for some time. This is due to the time it takes to close gaps between beads close to the interface that emerge upon the solitary wave hitting the interface, as shown in Figure \ref{fig3}.

After the initial wave hits the interface at time $t_1$, indicated in Figure \ref{fig4}, the emerging gaps lead to bead 49 freely travelling forward while bead 50 is travelling backwards. The beads at the interface are frozen in this state between times $t_2$ and $t_3$, when beads 49 and 50 exchange momenta in an elastic two-body collision. This new state is again frozen until time $t_4$, when bead 50 closes the gap to its right neighbour and induces the secondary transmitted wave. The reflected wave starts propagating away from the interface at time $t_5$ when bead 49 reaches its left neighbour. Similar delays have previously been reported for granular containers with mass interfaces \cite{VER02}.

\begin{figure*}\begin{center}
\includegraphics[width=0.25\textwidth]{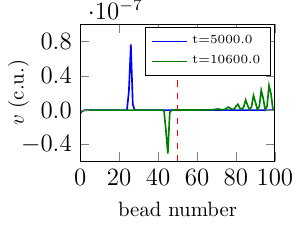}
\includegraphics[width=0.25\textwidth]{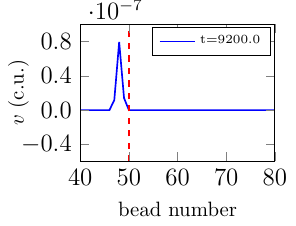}
\includegraphics[width=0.25\textwidth]{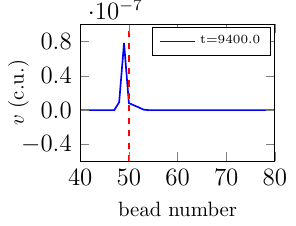}\\
\includegraphics[width=0.25\textwidth]{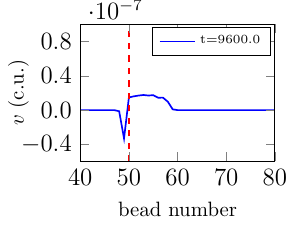}
\includegraphics[width=0.25\textwidth]{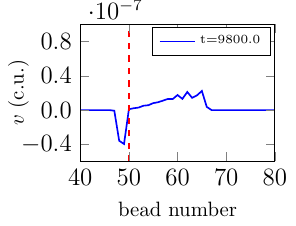}
\includegraphics[width=0.25\textwidth]{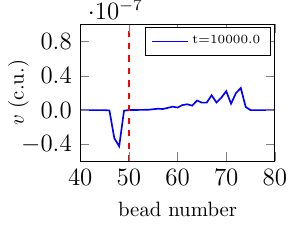}
\vspace{-0.2in}
\caption{Solitary wave propagation across an interface between media with $n_1=3.0$ and $n_2=2.5$.}
  \label{fig:large_small}
\end{center}\end{figure*}

Bead 50 travelling backwards and thereby inducing the small reflected wave is essentially an instance of chain fragmentation in the the right-hand side medium. When the initial solitary wave hits the interface at time $t_1$ a gap between bead 49 and bead 50 opens, and therefore the two media are disconnected. Bead 50 acts as a striker that hits the right medium inducing a solitary wave and chain fragmentation with some beads rebounding in the other direction.

Since the opening and closing of gaps is essential for the explanation of the secondary waves that emerge for interfaces with $\Delta n>0$, we do not expect them to be captured by any continuum model, as gaps are a feature of the discreteness of granular crystals.

\subsection{Decreasing interaction exponent}
\label{sec:DIE}

For a wave propagating across an interface with $\Delta n < 0$ the effects are quite different. Figure \ref{fig:large_small} shows the bead velocities before and after the initial solitary wave hits the interface for $n_1 = 3.0$ and $n_2=2.5$. There is a significant reflection and transmission, with the transmitted energy being converted into a multipulse structure. Unlike in the case discussed above, there is no delay of transmitted or reflected part of the wave.
At the moment when the reflected wave starts to propagate away from the interface, the transmitted energy forms a pulse with a nearly rectangular shape. In the case shown here this means the first 8--9 beads behind the interface have roughly the same velocity. This nearly rectangular pulse splits and goes on to form the transmitted multipulse structure.

In previous numerical simulations it was observed that for rectangular initial conditions, i.e. the first $l$ particles having the same initial speed, a multipulse structure with $l$ waves emerges \cite{NES02, HIN01}. The transmitted pulse in Figure \ref{fig:large_small} resembles such initial conditions.

\begin{figure}
\begin{center}
 		\includegraphics[width=0.35\textwidth]{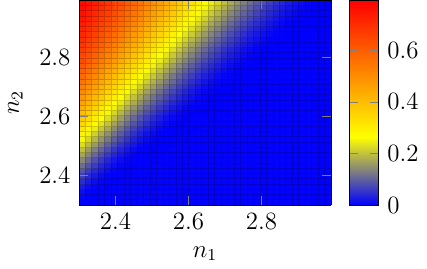}
 		\vspace{-0.2in}
		\caption{Reflected energy as function of $n_1$ and $n_2$.}
  \label{fig:reflected_energy_rel}
\end{center}\end{figure}

The reflected energy propagates away from the interface in the form of a single solitary wave, whose energy can be measured easily. We determine the range of beads that form the solitary wave by finding the beads with more than $0.1\% $ of the peak bead velocity in the wave. As the energy carried by the wave is not purely kinetic, we also need to consider the potential energy stored in the interaction between the beads. Then for a solitary wave which is made up of $s$ beads starting at the $k^\mathrm{th}$ bead, the wave energy is given by
\begin{equation}
  E = \sum_{i=k}^{k+s} m_i v_i^2 + \sum_{j=k}^{k+s-1} a \delta_{j,j+1}^{n_1} ,
\end{equation}
where the first and second sum are the kinetic and potential energy carried by the wave respectively. The fraction of reflected energy $E_\mathrm{refl}/E_\mathrm{in}$, where $E_\mathrm{refl}$ and $E_\mathrm{in}$ are the reflected and initial energy, is shown in Figure \ref{fig:reflected_energy_rel} as a function of $n_1$ and $n_2$.

\begin{figure}
\begin{center}
     \includegraphics[width=0.35\textwidth]{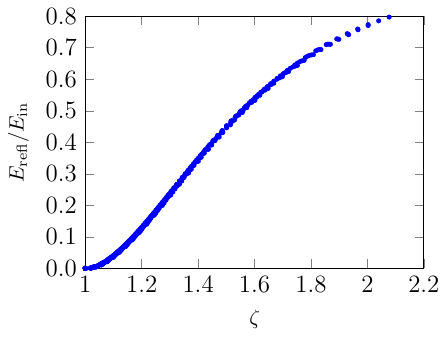}
    \vspace{-0.2in}
      \caption{Simulated reflected energy as a function of the new variable $\zeta$. }
  \label{fig:reflectedCollapsed}
\end{center}\end{figure}

For $\Delta n <0$, i.e. when significant reflection occurs, the contours in Figure \ref{fig:reflected_energy_rel} are linear and collapse around the point $(n_1, n_2) \approx (1.65, 1.65)$, which is outside the plotted domain. Therefore, the reflected energy can approximately be described in terms of a new variable
\begin{equation}
  \zeta = \frac{n_1 - 1.65}{n_2 - 1.65} ,
  \label{eqn:zeta}
\end{equation}
as shown in Figure \ref{fig:reflectedCollapsed}.

\subsection{Two-particle versus few-particle interactions}
\label{sec:TFI}
 
Let us consider an interface between media with different bead masses as discussed in \cite{VER01, VER02, SEN01, MAN02}, more specifically the case where a solitary wave is propagating from the lighter to the heavier medium. For large values of $n$ most of the kinetic energy of a propagating solitary wave is carried by a single bead. In this case we can approximate the dynamics to be completely described by at most two particles interacting with each other at a given point in time, as the velocities of all other particles are essentially zero.

Under these assumptions, the effects arising from wave propagation across the interface can be derived from the interaction between the last particle with the lighter mass $m$ at site $k$, and the first particle with the heavier mass $M$ at site $k+1$ \cite{MAN02}. For such a simple two-body collision, the velocities after the collision, $v'_k$ and $v'_{k+1}$, are completely determined by the conservation of momentum and energy without having to consider the details of the interaction potential: $v'_k = \frac{m-M}{m+M} v_k$, $v'_{k+1} = \frac{2m}{m+M} v_k$, where $v_k$ is the velocity of bead $k$ before the collision and bead $k+1$ is assumed to be initially at rest. Note that we need to assume the inter-particle force is repulsive and conservative, i.e. the collision is elastic and no energy is lost. As we have $m<M$, bead $k$ moves in the opposite direction causing the reflected wave, whereas bead $k+1$ moves to the right initialising the transmitted wave. A similar two-body argument can be given for the effects arising for a solitary wave propagating in the opposite direction \cite{MAN02}.

Returning back to our case of interfaces between regions with different interaction potential exponents $n$, we can consider the interactions under the same two-body approximation. However, as mentioned above, the details of the interaction are irrelevant for the outcome of a two body collision and since the bead masses are equal we get
$v'_k = 0 = v_{k+1}$, $v'_{k+1} =  v_k$, i.e. the two particles simply exchange momenta. Therefore, in contrast to the mass interfaces, here the reduction to a two-body problem cannot explain any of the observed interface effects. Therefore, all of the observed phenomena must be fundamentally connected to few-body interactions.

This can also be elucidated by recalling that the interface here is constituted by a bead with different interaction potentials on its left and right side, rather than the boundary between two beads, as is true for mass interfaces. Therefore, we need at least three beads to interact in order for the different interaction potentials to play a role in the wave propagation.

At interfaces with $\Delta n < 0$, the number of particles interacting at the interfaces is significantly larger than three. We can define the number of particles that interact at the interface as the number of excited particles just before the reflected and transmitted waves part. We observe that at the interface the first couple of beads in the right medium are collectively excited by the collision with the incoming solitary wave, as shown in Figure \ref{fig:large_small}. In the collision process these collectively excited beads in the right-hand side medium can be imagined as forming a quasi-particle with a mass equal to the sum of its constituent bead masses. The incoming wave accelerates several particles in the right medium simultaneously and is then reflected off this quasi-particle, due to its heavier mass.

The number of particles involved in the interaction at the interface might also be the explanation for the decrease in reflected energy when $n$ is increased uniformly on both sides. For larger $n$, the wavelength and therefore the number of particles involved in the formation of quasi-particles are smaller, which thereby could decrease the reflection.

\section{Periodicity in the granular chain}

\begin{figure}[h]
\includegraphics[width=0.95\linewidth]{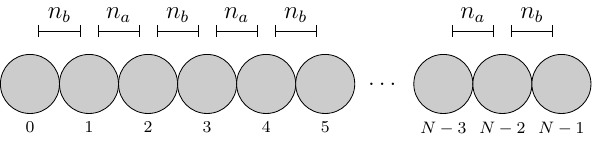}
\vspace{-0.1in}
\caption{Granular chain of spherical beads with alternating interaction potentials.}
\label{fig:periodic_chain}
\end{figure}

Another possibility for the implementation of the limit of many interfaces is to periodically alternate between two interaction exponents. Granular crystals with alternating bead masses, also called \textit{dimer} or \textit{di-atomic} granular crystals, show interesting wave propagation properties. Motivated by these results, we investigate the effects of periodically changing the interaction exponents in the chain. More specifically, in this section we study the wave propagation in granular crystals in which the interaction exponent is alternating between two values, $n_a$ and $n_b$, as depicted in Figure \ref{fig:periodic_chain}. We denote the difference between the two interaction exponents as $\text{d} n = n_b - n_a$.

\begin{figure}
\centering  
\includegraphics[width=0.35\textwidth]{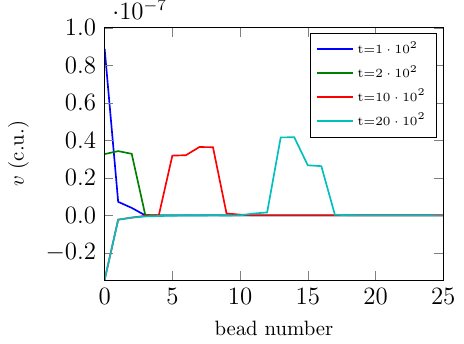}\\
\includegraphics[width=0.35\textwidth]{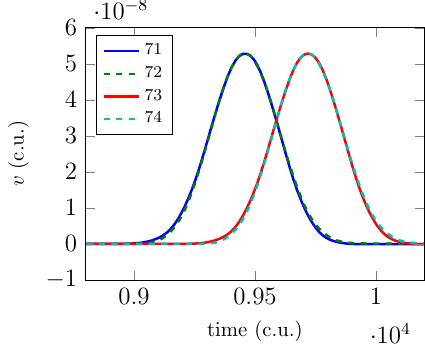}
\vspace{-0.2in}
      \caption{Solitary wave with pairwise collective behaviour for $n_a=2.5$ and $n_b=3.0$: emergence of solitary wave upon striking the chain from the left (top) and bead velocities for beads 71--74 in the chain (bottom). }
  \label{fig:periodic_pairs}
\end{figure}

We can distinguish three regimes of $\text{d} n$ with different qualitative behaviour. In the trivial case for $\text{d} n\approx 0$ we recover the behaviour for a monodisperse chain. In the limit $|\text{d} n|\to\infty$, with the constraint $n_a, n_b>2$, we find pairwise collective behaviour of the beads. Figure \ref{fig:periodic_pairs} shows the emergence of a solitary wave for a granular crystal with $n_a=2.5$ and $n_b=3.0$. A solitary wave is forming within the first 15 beads, however, neighbouring beads, which interact via a potential with the smaller exponent $n_a$, move in pairs. Figure \ref{fig:periodic_pairs} shows the bead velocity for four consecutive beads in the chain. The two bead pairs that interact with the exponent $n_a$ have practically identical velocities. Furthermore, for the qualitative behaviour of the emerging solitary wave it is irrelevant whether the chain starts with $n_a$ or $n_b$ at the left end. The only difference between these cases is the fraction of energy which is lost in the chain fragmentation, where the largest rebound velocity is higher when the chain starts with the smaller interaction exponent. 

\begin{figure}
\centering  
\includegraphics[width=0.35\textwidth]{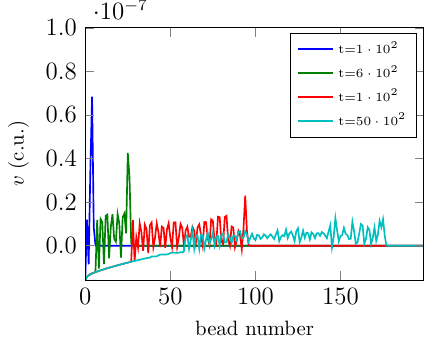}\\
\includegraphics[width=0.35\textwidth]{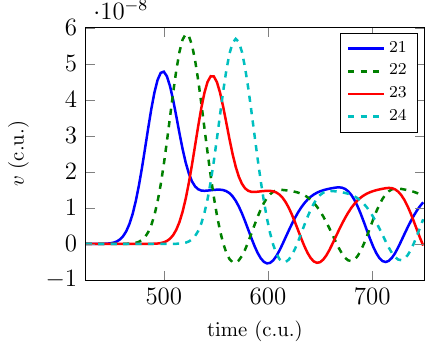}
\vspace{-0.2in}
    \caption{Wave decay without pairwise collective behaviour for $n_a=2.5$ and $n_b=2.555$: emergence of solitary wave upon striking the chain from the left (top) and bead velocities for beads 21--24 in the chain (bottom). }
  \label{fig:periodic_decay}
\end{figure}

Generally, for large exponent differences $\text{d} n$, the granular crystal behaves like a chain made up of particles with mass $2m$ interacting via a potential with exponent $\max(n_a,n_b)$. Hence, in this limit, the precise value of the smaller exponent becomes irrelevant.

There is an intermediate regime for $0<|\text{d} n|$, in which we find similar behaviour to chains with disorder in the interaction exponents. Figure \ref{fig:periodic_decay} shows the propagation of a wave for $n_a=2.5$ and $n_b=2.555$. The wave energy decays exponentially with the travelled distance, however the leading wave leaves a trail of smaller waves behind which is more regular than for chains with disorder in the interaction exponents. Neighbouring beads which interact via the smaller exponent do not behave collectively, as shown in Figure \ref{fig:periodic_decay}. However, we do observe that all beads, sufficiently far away from the left end, behave qualitatively similar to beads that have the same configuration of left and right interaction exponents. In other words, all even beads show qualitatively similar behaviour, and the same is true for all odd beads.

\begin{figure}
\centering  
\includegraphics[width=0.35\textwidth]{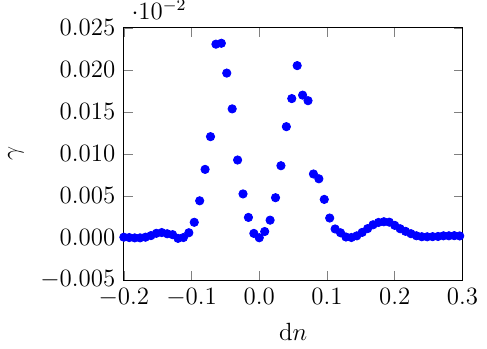}\\
\includegraphics[width=0.35\textwidth]{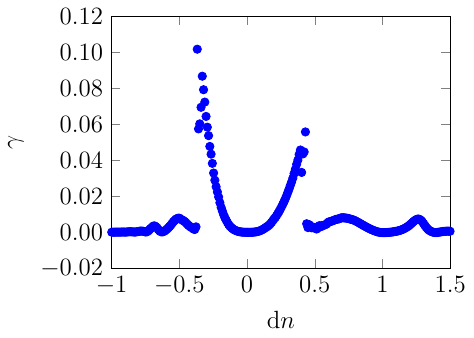}
\vspace{-0.2in}
      \caption{Decay rates for periodically arranged interaction exponents with $n_a=2.5$ (top) and $n_a=4.0$ (bottom). }
  \label{fig:periodicity_decay_rates}
\end{figure}

\begin{figure}
\centering  
\includegraphics[width=0.35\textwidth]{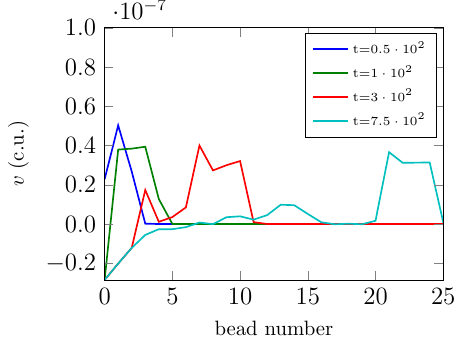}\\
\includegraphics[width=0.35\textwidth]{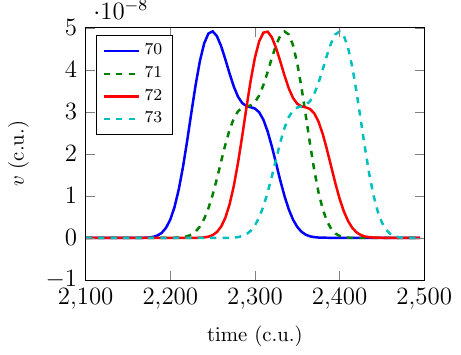}
\vspace{-0.2in}
  \caption{ Solitary wave with pairwise on resonance for $n_a=2.5$ and $n_b=2.635$: emergence of solitary wave upon striking the chain from the left (top) and bead velocities for beads 71--74 in the chain (bottom). }
  \label{fig:periodic_resonance}
\end{figure}

As for the analysis in Section \ref{sec:DIS}, we can determine the decay rate through a least-squares fit to the energy of the leading wave. The decay rate is shown for different $n_a$ in Figure \ref{fig:periodicity_decay_rates} as a function of $\text{d} n$. As expected the decay rate is zero for $\text{d} n=0$ and tends to zero for large $|\text{d}n|$. However, there are other local minima in addition to the local minimum at $\text{d} n=0$.

Figure \ref{fig:periodic_resonance} shows the wave propagation corresponding to the parameters first minimum for positive $\text{d} n$ in Figure \ref{fig:decay_rates}. Within around 25 beads we observe the formation of a stable compression wave and smaller separated secondary waves. The dynamics of the beads forming the wave more complex than for other waves we have discussed thus far. In contrast to waves forming in the limit of large $\text{d} n$, the wave emerging here is not of constant shape. Figure \ref{fig:periodic_resonance} shows the bead velocities for four consecutive beads in the chain. Beads 71 and 72 interact via a potential with the smaller exponent $n_a$, i.e. they correspond to beads that would behave collectively in the limit of large $|\text{d} n|$. We observe that the velocity of the two beads oscillates around their mean, during which the bead with the highest velocity changes three times. Hence, the peak of the wave periodically oscillates between the front and the back of the wave. These changes of the bead with the peak velocity correspond to several small collision-like interactions between beads that form the wave. These are not full collisions, as the beads in the wave do separate. For the parameters corresponding to the first minimum for negative $\text{d} n$ we observe similar behaviour.

\begin{figure}
\centering  
\includegraphics[width=0.35\textwidth]{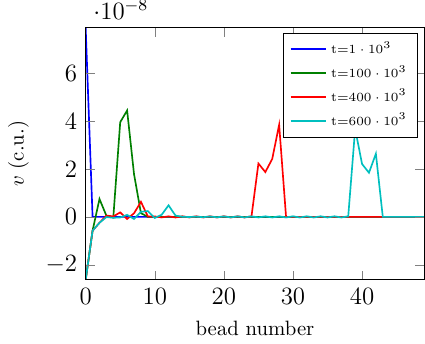}\\
\includegraphics[width=0.35\textwidth]{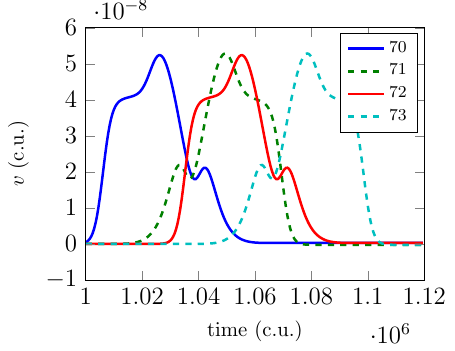}
\vspace{-0.2in}
    \caption{Solitary wave with pairwise on resonance for $n_a=4.0$ and $n_b=5.0$: emergence of solitary wave upon striking the chain from the left (top) and bead velocities for beads 71--74 in the chain (bottom). }
  \label{fig:periodic_resonance2}
\end{figure}

For $n_a=4.0$, there is one more minimum in the decay rate at both ends, as shown in Figure \ref{fig:periodicity_decay_rates}. The wave propagation for parameters corresponding to the second minimum are shown in Figure \ref{fig:periodic_resonance2}. The dynamics are similar to the first minimum, however the velocities of the  bead pairs oscillate five times around their mean while forming the wave.

The bead velocities shown in Figures \ref{fig:periodic_resonance} and \ref{fig:periodic_resonance2} show interesting symmetry properties. The velocity profiles of all odd and all even beads are qualitatively and quantitatively the same, only shifted in time. Furthermore, even though the bead pairs interacting via a potential with the smaller exponent do not move collectively, their velocity profiles show a mirror symmetry. This type of wave might not be considered a solitary wave, as the wave shape periodically changing. However, these waves are stable and spatially confined, as for classical solitary waves in granular crystals. We are not aware of any other reports on such stable waves with periodically changing wave shape for granular crystals.

\section{Local and global disorder in the granular chain via heterogeneous interaction exponents}
Disorder in diatomic granular chains was studied in \cite{ponson2010nonlinear} by viewing the granular chain as a spin chain composed of units that are each oriented in one of two possible ways. Recently, disorder in granular chains have been explored experimentally and through simulations by \cite{kim2018direct}, who realize the disorder by considering different densities and elastic moduli of the beads. In the present section, we shall explore an alternative route to disorder in the granular chain by considering arising from heterogeneity in the interaction exponents. 

\subsection{Local impurities in otherwise homogeneous chains}
\begin{figure}[h]
\begin{center}
\includegraphics[width=0.45\textwidth]{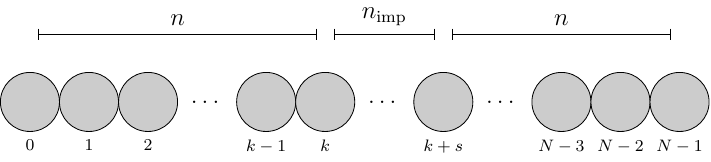}
\vspace{-0.1in}
\caption{Granular chain with interaction exponent $n$ with an impurity of $s$ neighbour interactions with exponent $n_\mathrm{imp}$. }
\label{fig:impurity}
\end{center}\end{figure}%

Knowing that the abrupt changes in the interaction exponent between two media can cause complex wave propagation phenomena, one might ask what happens when there is a change in the interaction exponent in a localised region within an otherwise monodisperse chain; see Figure \ref{fig:impurity}. We assume that the force between $s$ consecutive bead-bead pairs within the chain is governed by the interaction exponent $n_\mathrm{imp}$ which are embedded in a chain with interaction exponent $n$. In this set-up, $s+1$ beads have at least one interaction with a neighbouring bead with the exponent $n_\mathrm{imp}$. In order to distinguish impurities from interfaces, we denote the difference in the exponent between the medium and the impurity as $\delta n = n_\mathrm{imp} - n$.

\begin{figure}
\begin{center}
    \includegraphics[width=0.35\textwidth]{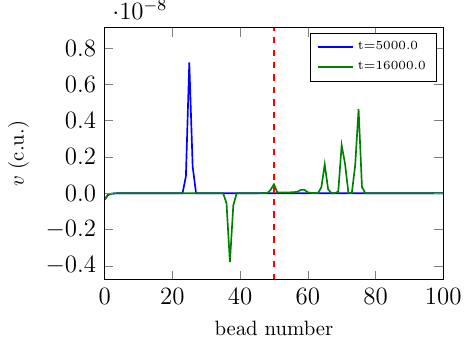}\\
    \includegraphics[width=0.35\textwidth]{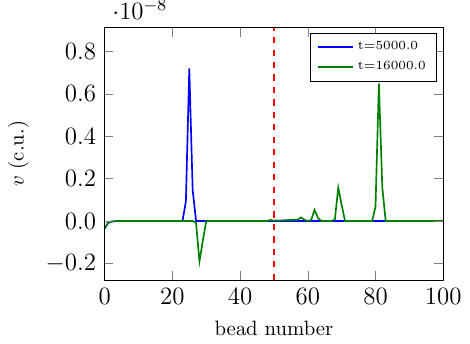}
   \vspace{-0.2in}
      \caption{Granular crystals with impurities with $s=1$, $n=3.0$ for $n_\mathrm{imp}=3.5$ (top) and $n_\mathrm{imp}=2.5$ (bottom). The red dashed line indicates bead $k$ (c.f. Figure \ref{fig:impurity}). }
\label{fig:impurity_example}
\end{center}\end{figure}

Figure \ref{fig:impurity_example} shows two examples of granular crystals with impurities for $\delta n > 0$ and $\delta n < 0$. Upon hitting the impurity, the initial solitary wave is partially reflected with the remaining energy being converted into a multipulse structure. In contrast to interfaces, here we obtain similar qualitative behaviour independently of the sign of $\delta n$.

\begin{figure}
\begin{center}
  \includegraphics[width=0.35\textwidth]{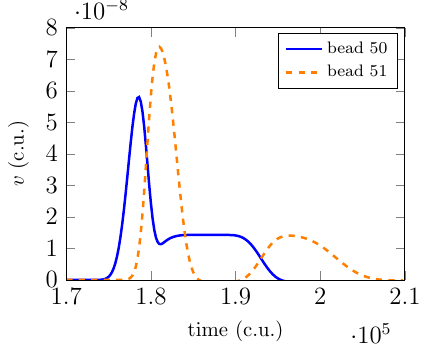}\\
\includegraphics[width=0.35\textwidth]{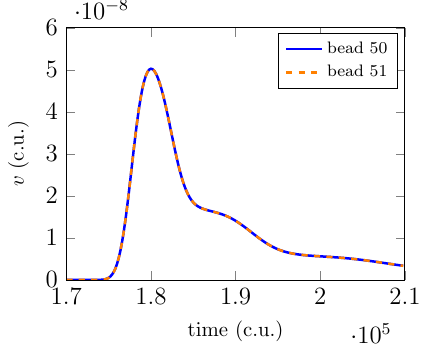}
\vspace{-0.2in}
   \caption{Impurity bead velocities with $n=4.0$ and $s=1$ for $\delta n=-0.25$ (top) and $\delta n= -1.5$ (bottom). The two lines are not distinguishable in the bottom panel.}
\label{fig:bead_pairs}
\end{center}\end{figure}%

\begin{figure}
\begin{center}
\includegraphics[width=0.35\textwidth]{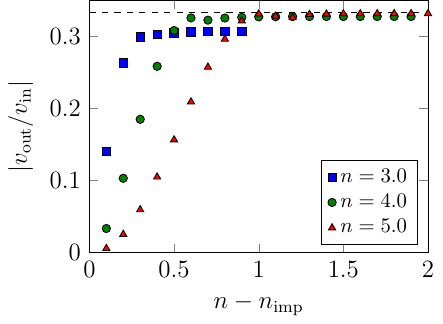}
\includegraphics[width=0.35\textwidth]{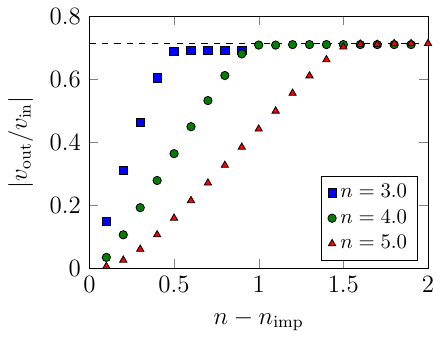}
\vspace{-0.2in}
    \caption{Ratio of the peak bead velocities between reflected and incoming solitary wave for impurities with different numbers of impurities $s$: $s=1$ (top) or $s=5$ (bottom). The dashed line indicates the velocity of a bead with mass $m$ scattering of a bead with mass $(s+1)m$. }
\label{fig:sparticlereflection}
\end{center}\end{figure}%

An interesting observation we make is that for large negative $\delta n$, the impurity beads behave collectively, i.e. their velocities are identical. Figure \ref{fig:bead_pairs} shows two examples for the impurity bead velocities during the interaction with the incoming solitary wave for $s=1$. Whereas for $\delta n=-0.25$ there is an asymmetry between the two impurity beads, they are completely synchronised for $\delta n = -1.5$. This clearly shows quasi-particle behaviour in which the two impurity beads act as one particle of mass $2m$, similar to the observations for interfaces.

We can compare the collision of the solitary wave with the impurity to the two-body approximation as discussed in Section \ref{sec:TFI}. Figure \ref{fig:sparticlereflection} shows the the ratio of the peak bead velocities between reflected and incoming solitary wave $|v_\mathrm{out}/v_\mathrm{in}|$ for different values of $n$ and $s$. For large interaction exponent differences between the medium and the impurity, the reflected velocity converges  to the velocity one would get from a particle with mass $m$ colliding with a particle of mass $(s+1)m$ at rest. Therefore the process of the solitary wave interacting with an impurity is described well by the quasi-particle explanation. For small values of $n$ the reflected velocity does not quite reach the value predicted by the two-body approximation. This is due to the fact that the incoming wave has a larger wavelength, hence approximating it by the motion of a single bead is inaccurate.

\subsection{Disorder in the granular chain}
\label{sec:DIS}
We now consider the case where the disorder is not localized, but rather occurs over the length of the granular chain, resulting in a fundamentally heterogeneous chain structure. First, recall that solitary waves propagating through a granular chain with randomised masses experience exponential decay were previously studied \cite{MAN02}. In \cite{MAN02} the masses are set to
\begin{equation}
  m_i = \bar{m}\left( 1 + r_i \sigma\right), \qquad i\in\{ 0, \dots, N-1 \} ,
  \label{eqn:m_random}
\end{equation}
with $\bar{m}$ being the mean mass, $r_i$ being a random number that is uniformly distributed in $[-1,1]$, and $\sigma$ a parameter weighting the randomness of the masses. It could be observed that the solitary wave energy $E$ decays as 
\begin{equation}
  E(x) = E_0 \exp(-\alpha_E x) ,
\end{equation}
where $E_0$ is the initial energy, $x$ is the position in the chain and $\alpha_E$ is the decay rate. Furthermore, it was found that $\alpha_E \propto \sigma^2$ for small enough $\sigma$.

We shall investigate chains with randomised interaction potential exponents where
\begin{equation}
  n_i = \bar{n}+ r_i \epsilon, \qquad i\in\{ 0, \dots, N-1 \} ,
  \label{eqn:n_random}
\end{equation}
with the mean exponent $\bar{n}$, $\epsilon>0$ and $r_i$ defined as for \eqref{eqn:m_random}. Whereas the scenario studied in \cite{MAN02} corresponds to a chain of beads with randomised sizes, the case we consider here corresponds to a chain with randomly varying contact surface geometries between the beads.

\begin{figure}[h]
\includegraphics[width=0.95\linewidth]{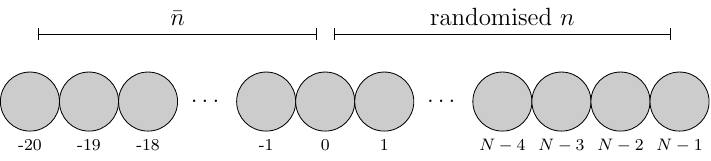}
\vspace{-0.1in}
\caption{Set up of the granular chain for the simulation of solitary wave propagation under randomised $n$. The first 20 beads constitute a monodisperse chain and the randomisation starts from bead 0. In the simulations $N=1000$ was used.}
\label{fig:n_random}
\end{figure}

In order to avoid chain fragmentation effects in the simulation, the chain is set up as shown in Figure \ref{fig:n_random} with a monodisperse chain of 20 beads at the beginning of the chain. In this set-up, the solitary wave can develop before it reaches the part of the chain with randomised $n$, so that its energy can be measured before the decay.

\begin{figure}
\centering  
\includegraphics[width=0.35\textwidth]{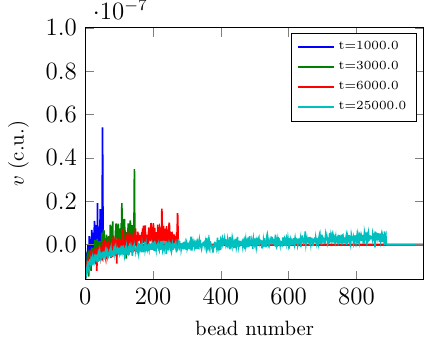}\\
\includegraphics[width=0.35\textwidth]{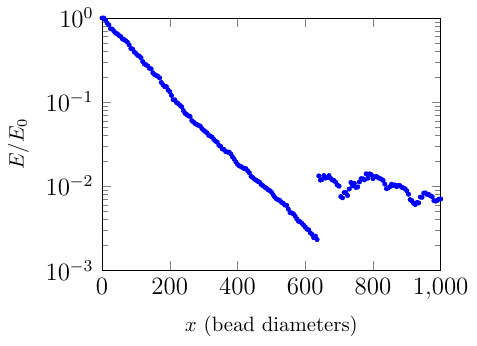}
\vspace{-0.2in}
  \caption{Decay of the solitary wave energy in a chain with disorder in the interaction exponents for $\bar{n}=2.5$ and $\epsilon=0.04$. \label{fig:decay}}
\end{figure}

\begin{figure}
\centering
\includegraphics[width=0.35\textwidth]{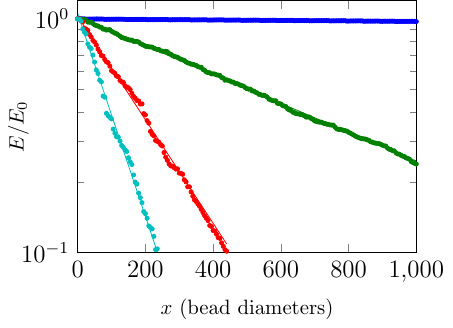}\\
\includegraphics[width=0.35\textwidth]{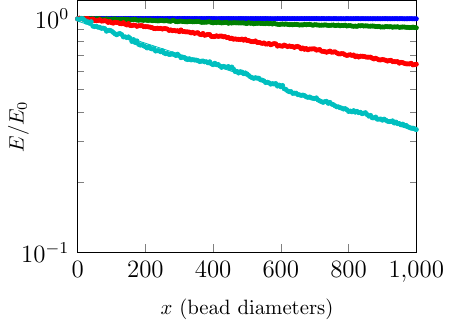}
\vspace{-0.2in}
\caption{Wave decay in chain disorder in the interaction potential exponents for average interaction potential values (top) $\bar{n} = 2.4$ and (bottom) $\bar{n}=3.2$, and with randomness parameters $\epsilon$: $0.002$ (\crule[blue]{0.5em}{0.5em}), $0.014$ (\crule[green!50.0!black]{0.5em}{0.5em}), $0.026$ (\crule[red]{0.5em}{0.5em}) and $0.04$ (\crule[cyan]{0.5em}{0.5em}). The travel distance is given in units of the bead diameter. The dots represent the simulation data and the solid lines are fitted exponentials.}
\label{fig:exponential_decay}
\end{figure}

The propagation of the solitary wave through a medium with randomised interaction potentials is shown in Figure \ref{fig:decay}(top). As for chains with disordered masses \cite{MAN02}, the initial solitary wave leaves a trace of fluctuating bead velocities behind, to which it continuously looses energy as it traverses the chain. We can quantify the decay of the solitary wave by measuring the total kinetic and potential energy it carries, and the energy of the solitary wave relative to the initial energy $E/E_0$ is shown in Figures \ref{fig:decay}(bottom) and \ref{fig:exponential_decay}. Furthermore, as we have uniform bead masses and the energy is conserved in the system, the bead velocities in the fluctuations right behind the leading wave are smaller than than its maximum bead velocity.

Due to the non-linear nature of waves in granular chains, the trailing fluctuations are slower and do not interact with the leading wave once they have separated. However, once the leading wave has lost enough energy it becomes smaller than fluctuations that have earlier been induced by the leading wave. Therefore, these earlier fluctuations can overtake the leading wave. Hence, for long enough propagation distances, the leading wave is indistinguishable from the induced fluctuations, at which point one can say that the initial solitary wave has fully disintegrated. The jump in the energy of the leading pulse in Figure \ref{fig:decay} shows the point at which the initial solitary wave is overtaken by fluctuations.

In the framework of shock attenuating materials, this kind of wave disintegration may be favourable over shock attenuators consisting of composite materials. Unlike attenuators using interfaces in composite materials, the energy that is extracted from the wave is almost uniformly distributed over all degrees of freedom in the chain. This reduces the maximal single impact at the right end of the chain for fully disintegrated waves.

We observe that the induced fluctuations depend on the size of the leading solitary wave, i.e. larger waves cause larger fluctuations. This translates to the energy loss depending on the energy of the leading wave. Figure \ref{fig:exponential_decay} shows the energy of the leading wave as a function of distance travelled in the chain. We can conclude that the energy loss per unit distance travelled is proportional to the wave energy as it decays exponentially.

The straight lines in Figure \ref{fig:exponential_decay} are the result of a least-squares fit of an exponential decay function
\begin{equation}
  f(x) = a e^{- \gamma x} ,
\end{equation}
to the wave energy, where the wave energy is given relative to the initial wave energy and the travel distance $x$ is given in units of bead diameters. In our units the decay rate $\gamma$ is the decay of the wave energy per traversed bead. The deviations from the exponential decay come from the randomness of the interaction exponent in the chain, which causes inhomogeneities in the $n$-differences between beads along the propagation direction.

\begin{figure}
\centering
\includegraphics[width=0.40\textwidth]{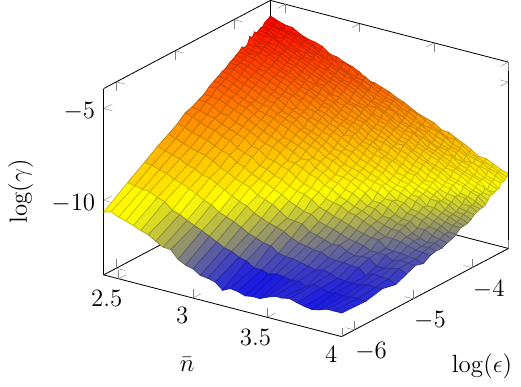}
\vspace{-0.2in}
\caption{Decay rate $\gamma$ of the solitary wave as function of $\bar{n}$ and $\epsilon$.}
\label{fig:decay_rates}
\end{figure}

The decay rate as a function of the randomness $\epsilon$ and $\bar{n}$ is shown in Figure \ref{fig:decay_rates}. We observe the following relations
\begin{equation}
  \log(\gamma)\propto\log(\epsilon), \qquad \log(\gamma)\propto \bar{n} ,
\end{equation}
from which we infer the ansatz for a logarithmic model for $\gamma(\epsilon, n)$ as
\begin{equation}
  \log(\gamma) = \alpha \log(\epsilon) + \tilde{B} \bar{n} + \tilde{C} ,
  \label{eqn:decay_rate_fit}
\end{equation}
where $\alpha$, $\tilde{B}$ and $\tilde{C}$ are parameters which are determined by a least-squares fit. For the fit we are only using data with $\log(\gamma)>-10$, as for very small decay rates the obtained data is unreliable due to the finite precision of the exponential decay fit through which they are computed. The results for the fit parameters are shown in Table \ref{tab:decay_rate_fit}. 

\begin{table}[htb]
	\centering
	\begin{tabular}{@{}l l l@{}}
	\hline
	Parameter & Value & $\sqrt{\sigma}$ \\
	\hline
	$\alpha$ & $2.08453172$ &  $0.0079$ \\
	$\tilde{B}$ & $-3.51075448$ &  $0.014$ \\
	$\tilde{C}$ & $10.94456937$ &  $0.077$ \\

	\hline
	\end{tabular}
	\caption{Fit parameters of the fit function in Equation \eqref{eqn:decay_rate_fit}. Here $\sigma$ is the variance and $\sqrt{\sigma}$ is the standard deviation.}
	\label{tab:decay_rate_fit}
\end{table}

We can rearrange Equation \eqref{eqn:decay_rate_fit} to obtain
\begin{equation}
  \gamma(\epsilon, \bar{n}) = C \epsilon^\alpha \beta^{\bar{n}}
  \label{eqn:decay_rate} ,
\end{equation}
with $\beta=\exp(\tilde{B})=0.030$ and $C=\exp(\tilde{C})=56646$. Hence the wave energy decays as
\begin{equation}
  E(x) \propto \exp\left( - C \epsilon^\alpha \beta^{\bar{n}} x \right).
\end{equation}
Interestingly, we find the same relation between the randomness parameter and the decay rate as in the case for disorder in the bead masses (note that the different definitions of \eqref{eqn:m_random} and \eqref{eqn:n_random} do not affect the exponent of $\epsilon$ in the decay rate), as $\gamma \propto \epsilon^\alpha$ with $\alpha\approx 2$.

For the parameters tested here the maximal decay rate is obtained at $\bar{n}=2.4$, $\epsilon=0.04$ for which the decay rate \eqref{eqn:decay_rate} is $\gamma \approx 0.02$. At this decay rate the wave energy is halved approximately every $15$ beads.

Another question one might ask is how the differences in interaction potential influence the energy transmission in the granular chain. As shown above, the mechanisms that lead to wave reflection at interfaces are few-particle interactions, however, the media in the case of disorder in the chain are reduced to a single particle on each side of the interface. Therefore, the knowledge about the effect of differences in the interaction potential exponent cannot necessarily be transferred to the case of disorder.

\begin{figure}
\centering  
\includegraphics[width=0.35\textwidth]{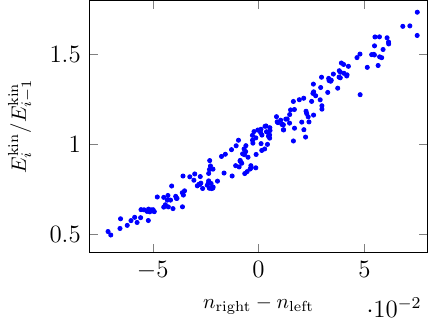}\\
\includegraphics[width=0.35\textwidth]{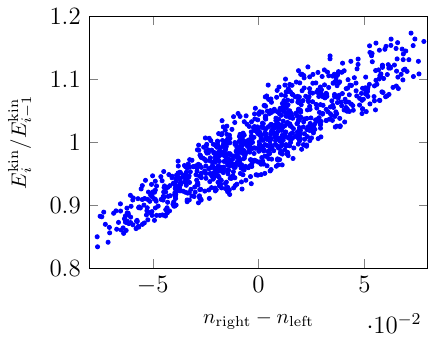}
\vspace{-0.2in}
      \caption{Energy transfer from bead $i-1$ to bead $i$ measured by the ratio of maximum kinetic energies $E^\mathrm{kin}_i/E^\mathrm{kin}_{i-1}$. We take $\bar{n}=2.4$ (top) and $\bar{n}=3.7$ (bottom), fixing $\epsilon = 0.04$.}
  \label{fig:transfer}
\end{figure}

As a measure of the transmitted energy from bead $i-1$ to bead $i$, we compare their maximal kinetic energy
\begin{equation}
  E^\mathrm{kin}_i = \frac{m v_i^2}{2} .
\end{equation}%
As shown in Figure \ref{fig:transfer}, the difference $n_\mathrm{right}-n_\mathrm{left}$ between the exponents in the interaction with the right and left neighbour of bead $i$ is a good parameter to describe how much kinetic energy is transmitted to it from bead $i-1$. On average, energy is reflected for $n_\mathrm{right}<n_\mathrm{left}$, which resembles the case of interfaces between media with different interaction exponents with $\Delta n<0$. For $n_\mathrm{right}>n_\mathrm{left}$ bead $i$ reaches an even higher maximum kinetic energy. This can be understood by considering the shapes of the interparticle forces as a function of the bead overlap. For larger interaction exponents, the force is smaller for small overlaps and increasing more steeply as the overlap is increased. Therefore, early on in the acceleration process, the $i^\mathrm{th}$ bead is accelerated by bead $i-1$ without transferring much energy to bead $i+1$. In other words, the $i^\mathrm{th}$ bead has more time to be accelerated before it has to pass on the energy to bead $i+1$. The variance in the energy transfer is in accordance with our findings that processes at interfaces between media with different interaction exponents are few-particle processes. Therefore, we expect the energy transfer to not only depend on $n_\mathrm{left}$ and $n_\mathrm{right}$, but also on interactions with other beads close by.

It is remarkable that even though the phenomena at mass and interaction exponent interfaces are due to different processes, we find the same qualitative behaviour for disorder, which can be seen as the limit of many interfaces. Moreover, we find the same dependence of the decay rate on the randomness parameter. However, in contrast to disorder in the bead masses, we find an additional dependence on the average interaction exponent $\bar{n}$. This additional dependence can be explained by considering the scaling properties of the equation of motion \eqref{eqn:eom2}. Without externally applied forces, changing the overall mass is equivalent to changing the overall time scale of the dynamics, as we can remove any changes in the overall mass by appropriate re-scaling of time. This is not the case for changes of the average interaction exponents.

In terms of real world applications and experiments, although the interpretation of small random differences in masses is clear, the meaning behind small random differences in interaction exponents is less so. However, recall that the interaction potential depends on the geometry of the contact region \cite{SEN01}, and hence when assuming a power-law potential, the value of the exponent $n_{i,i+1}$ is determined by the contact region geometry \cite{JOH01, SEN01}. As an example, since $\overline{n}=2.5$ corresponds to the contact geometry between two spheres, one can see a value of $n=2.5+\epsilon$ as corresponding to the contact geometry between two objects which take the form of slightly deformed spheres (provided that $|\epsilon | \ll 1$ is sufficiently small). Therefore, given a granular chair composed of slightly irregular spheres in a random arrangement, the choice of $n=2.5+\epsilon$ is a sensible modeling assumption. More generally, for a ``perfect" contact geometry corresponding to $n = \overline{n}$, is is sensible to consider $n=\overline{n}+\epsilon$ as accounting for small defects irregularities in contact geometry from the baseline ``perfect" geometry. With that said, a remark is in order. While the effective interaction potential exponent $n$ may be changed in experiments by altering the contact geometry between the beads; note that changing the contact geometry will also change the prefactors $a$ \cite{JOH01, LAN01}. We ignore this latter issue in the present theoretical work, but remark that a corresponding change in prefactors, so that $a_{i,i+1}=a(\epsilon)$, may be needed for the most accurate comparison of theoretical simulations with real experiments. 

\section{Composite mass - interaction exponent interfaces}
\begin{figure}[h]
\begin{center}
\includegraphics[width=0.45\textwidth]{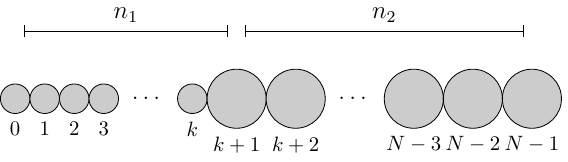}
\vspace{-0.1in}
\caption{Combined mass and interaction exponent interface. Note that the mass on the right-hand side can also be smaller than on the left.}
\label{fig:NMinterface}
\end{center}\end{figure}

Now that we know about the effects of both mass (from other works) and interaction exponent interfaces, we should comment on the combination of both types of interfaces influences the solitary wave propagation (for a schematic, see Figure \ref{fig:NMinterface}). In order to keep the parameter space manageable, we fix the parameters of the left medium to $n_1=3.0$ and bead mass $m_1=m$ as used in the previous simulations, and only change the parameters for the right medium. We vary $\Delta n$ in  $[-0.5,0.5]$ and the ratio $m_2/m_1$ between the bead mass on the left and on the right in $[1/8, 8]$.

\begin{figure}
\begin{center}
\includegraphics[width=0.40\textwidth]{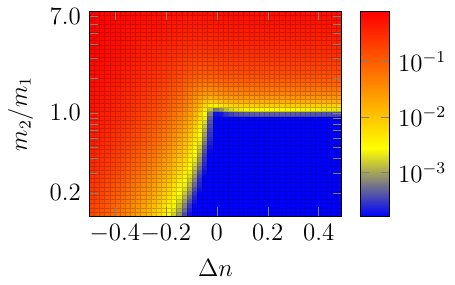}
\vspace{-0.2in}
\caption{Reflected energy as function of $\Delta n$ and $m_1/m_2$.}
  \label{fig:reflected_energy_relNM}
\end{center}\end{figure}

Figure \ref{fig:reflected_energy_relNM} shows the reflected energy as a function of the two interface parameters. Note that for visualisation purposes where no reflection was detected the value for the reflected energy was set to the minimum value detected over the entire parameter range. Therefore, for the entire dark blue area in the plot, no significant reflection could be detected.

In the bottom-right part of Figure \ref{fig:reflected_energy_relNM} there is no reflection; this is what we would expect from what we know of both types of interfaces in this parameter regime. Similarly, for the parameters in the top left corner both types of interfaces separately show reflection, which is also the case for the combined interface. In the top-right corner, i.e. for $\Delta n>0$ and $m_2/m_1>1$, the reflection is mainly determined by the mass difference with significant reflection for all $m_2/m_1>1$. This seems plausible as the difference in interaction exponents for $\Delta n>0$ had only small effects on the wave propagation.

However, for $\Delta n<0$ and $m_2/m_1<1$, we arrive at contradicting predictions using our knowledge from the separate study of these two types of interfaces: for mass interfaces we expect a full disintegration of the wave without any reflection, but for interaction exponent interfaces we would predict the emergence of a multipulse structure with significant reflection. For the combined interface we observe a transition between regions with and without reflection. The contour at which the transition occurs can be interpreted in the framework of the quasi-particle explanation: it is the set of parameters where the effective mass of the quasi-particle on the right-hand side and the bead mass on the left-hand side are equal. In other words, the mass difference in the quasi-particle collision is balanced out by the difference of the bead masses $m_1$ and $m_2$.

\section{Discussion}
We have investigated interfaces between distinct Hertz-like granular crystals with different interaction exponents $n_1$ and $n_2$. For $\Delta n>0$, we obtain mainly transmission of the incoming wave with one transmitted and one reflected secondary wave forming at the interface with a time delay to the primary transmitted wave. The secondary waves are around one order of magnitude smaller in peak bead velocity than the original wave and their formation involves the opening and closing of gaps in the vicinity of the interface. The emergence of these secondary waves is thus a direct consequence of the discreteness of the granular chain. This effect is not expected to be captured by any models using the continuum approximation.

For interfaces with $\Delta n<0$, both transmission and reflection of the solitary wave occur. Whereas the reflected part of the initial energy propagates away from the interface in form of a single solitary wave, the transmitted energy is converted into a multipulse structure. Combining both mass and interaction exponent interfaces we could find a transition between parameter regimes with and without reflection of the solitary wave. This is in accordance with our explanation for reflection in the absence of bead mass differences for the case of pure interaction exponent interfaces.

Although similarities between the effects at mass and interaction exponent interfaces exist, the underlying dynamics at the vicinity of the interface are fundamentally different. Whereas mass interfaces can be explained in terms of collisions between single beads with different masses, the phenomena at interaction exponent interfaces with $\Delta n<0$ involve few-body dynamics close to the interface. Moreover, we found that the first few beads on the right-hand side of the interface are collectively accelerated causing the partial reflection of the incoming wave. We were able to give a simple interpretation of the process in terms of a quasi-particle collision.
 
We have also considered the effect of disorder and periodicity in the interaction exponents in Hertz-like potentials on the propagation of solitary waves along Hertz-like granular chains. For randomised interaction potentials, the energy of the waves was found to decay exponentially in space by inducing noise in the bead velocities to which the energy is lost. We were able to find a law for the decay rate in terms of the mean interaction exponent and the randomness parameter. The decay rate scales in the same way with the randomness parameter as for disorder in the bead masses. As expected, great variations in bead contact geometry will result in more rapid degradation of the solitary wave, while more homogeneous beads, with only very small defects (as might be expected in experiments) will allow solitary wave propagation for a reasonable time interval before degradation of the wave is very noticeable. As commented on earlier, these results are akin to those of \cite{MAN02} for disorder in the bead masses.
 
In granular chains with interaction exponents alternating between two fixed values, we have identified distinct parameter regimes giving qualitatively different wave propagation properties. For large differences between the interaction exponents, we find pairwise collective behaviour of the beads, where neighbouring beads interacting via an interaction with the smaller exponent move as pairs, forming a solitary wave. Recalling the geometric meaning of the interaction exponents, this suggests that it is possible to obtain solitary waves, akin to those found for beads of one uniform contact geometry, in the case where contact geometries differ. This suggests that repetitive, periodic impurities can still permit solitary waves, at least in the regimes where differences between interaction exponents for neighbouring beads are large. 

In contrast, for intermediate values of the periodic interaction exponent differences, we find mainly exponential decay of the wave energy, similar to what we found in the case of randomised interaction exponents modelling disorder. However, there are certain parameters for which a different type of stable, confined waves develop, even though the beads do not move in pairs. These waves are different from classical solitary waves in granular chains, as their wave shape is not constant, but changes periodically. It is possible that these are the analogue ``disorder" solution to discrete breather solutions seen in uniform ``ordered" granular chains with one or few impurities \cite{br1}. Discrete breathers have been observed in other types of periodic media, including diatomic chains \cite{br2,br3}, therefore their appearance in media which is periodic due to a periodic contact geometry change is not wholly unexpected. 

One extension of this work might be the study of interfaces involving different types of interaction potentials. In this paper, we have considered interactions that are purely based on contact forces, which lead to Hertz-like power-law potentials. However, one could imagine having additional or different types of interactions between the particles. We are aware of one study using an external magnetic dipole field to introduce additional forces between the beads due to magnetisation. The additional interaction introduces dispersion, which in turn leads to wave attenuation \cite{LEN01}. These different types of interactions could be used to create interfaces which yield interesting wave propagation dynamics.

Another logical step in the study of interfaces between granular crystals with different interaction exponents under contact forces is the generalisation to higher dimensions. In higher dimensions the packing of the beads and the direction of the compression wave with respect to the main axes of the granular crystal influence the propagation in monodisperse media \cite{MNJ01}. When interfaces are present, higher dimensions offer additional degrees of freedom for the wave propagation, namely the angle at which the wave meets the interface. In 2D a law for the refraction and reflection of solitary waves at mass interfaces similar to Snell's law in optics was found \cite{TIC01}. As we have shown, in 1D there are similarities between the effects arising when solitary waves are propagation across mass and interaction exponent interfaces. Therefore, in 2D there could be similar refraction and transmission effects for interaction exponent interfaces. 2D hexagonal crystals have also been studied \cite{leonard2014traveling}, with wave propagation across interfaces formed by beads of different material considered recently in \cite{2dhex}. It may be interesting to extend such studies to materials formed with different interaction exponents.

\appendix
\section{Numerical approach and error analysis}
For a given chain with $N$ beads Equation \eqref{eqn:eom2} constitutes a coupled set of $N$ second order non-linear ODEs. Defining the vector of bead displacements as 
\begin{equation}
  \vec{u} = [u_1,u_2,\dots ,u_N]^T,
\end{equation}
we can formally write the equation of motion \eqref{eqn:eom} as
\begin{equation}
  \ddot{\vec{u}} = \vec{f}(\vec{u}) ,
  \label{eqn:eomk}
\end{equation}
where $\ddot{\vec{u}}$ denotes the second time derivative of $\vec{u}$. By defining the combined vector of bead displacements and velocities
\begin{equation}
  \vec{z} = [u_1, \dots, u_N, v_1, \dots , v_N]^T = \begin{bmatrix}
  \vec{u}\\
  \vec{v}
  \end{bmatrix}
\end{equation}
where $\vec{v} = [v_1, \dots, v_N]^T$, Equation \eqref{eqn:eomk} is converted into a system of 2N first order non-linear ODEs that can be written as
\begin{equation}
  \dot{\vec{z}} = \vec{F}(\vec{z}) =
  \begin{bmatrix}
    \vec{v} \\
 \vec{f}(\vec{u})
  \end{bmatrix} .
  \label{eqn:eom_vec}
\end{equation}
The main computational step for simulating the dynamics in granular crystals is solving Equation \eqref{eqn:eom_vec}, for $N$ of order $10^2$ to $10^3$.

All of the code written for carrying out the simulations and data analysis is written in Python, where we mostly use arrays from the Python extension NumPy for data storage. Where applicable, we use the scientific computing library SciPy which has implemented functionality similar to MATLAB.

As mentioned above, the equation of motion for the particles at the boundaries has only one interaction term since it has only one neighbour. In order to avoid having to use separate equations for the boundaries or looping through particles in the ODE solver, we added two ghost particles at the ends of the chain with the interaction prefactor $a$ to the first real particle being zero. Therefore the state vector $\vec{z}$ in the simulation becomes
\begin{equation}
  \vec{z}_\mathrm{sim} = [0, \vec{u}, 0, \vec{v}]^T ,
\end{equation}
where there are no extra entries for the velocity of the ghost particle, since they do not move. This way we can write the first order ODE \eqref{eqn:eom_vec} solely in terms of Numpy-array objects.

For solving the ODE system \eqref{eqn:eom_vec} we use a solver provided by the built-in ODE interface \textit{scipy.integrate.ode} in the SciPy library. We use the variable-coefficient ODE solver VODE, which in turn uses Adams-Moulten and backward differentiation methods depending on the stiffness of the system.

As there are no analytical solutions of the full discrete ODE system \eqref{eqn:eom_vec}, there is no direct way to determine the correctness and accuracy of our numerical scheme used for our simulations. However, we can make use of of both previous results and expected physical properties of the system to evaluate the numerical accuracy. Simulations of wave propagation across mass interfaces are produced by our numerical scheme and can be used to correctly emulate the behaviour in the original publication \cite{VER01}. Similarly our simulations show good qualitative agreement with results in the literature in the case of randomised bead masses and solitary wave collisions \cite{SEN01,MAN01}.

In order to evaluate the accuracy quantitatively we can make use of the energy conservation in the chain. As there are no dissipative effects built in to the model used to describe the dynamics, we expect all bead-bead collisions to be fully elastic, i.e. conserve the total energy. Therefore the energy loss---or gain---qualifies as a measure of the accuracy of the simulation. We use the total kinetic energy
\begin{equation}
  E^\mathrm{kin} = \sum_{i=0}^{N-1} m_i v_i^2 ,
\end{equation}
to determine how well the energy is conserved. We define the energy conservation error as
\begin{equation}
  \mathrm{error} = \left| \frac{E^\mathrm{kin}_\mathrm{start} - E^\mathrm{kin}_\mathrm{end}}{E^\mathrm{kin}_\mathrm{start}} \right| ,
\end{equation}
where $E^\mathrm{kin}_\mathrm{start}$ and $E^\mathrm{kin}_\mathrm{end}$ are the total kinetic energies in the chain at the start and the end of the simulation respectively.

Every granular crystal without externally applied forces reaches a steady-state of the bead velocities for long enough simulation times, i.e. the beads move at constant velocity with every bead having a smaller velocity than its right-hand-side neighbour. In terms of bead velocities the steady state translates to the bead velocities increasing  monotonously with the bead number%
Two subtleties need to be pointed out here: firstly, in order to have purely kinetic energy in the system we additionally need to require the bead overlap to be zero and secondly, for all practical purposes it is enough to require
\begin{equation*}
  v_{i+1}-v_i > -10^{-5}\times \max_{j \in \{ 0, \dots, N-1\}}  |v_j| ,
\end{equation*}
rather than actual monotonicity. This simplification reduces the simulation time needed to reach the steady state by orders of magnitude, while the additional errors introduced by it are negligible compared to the overall energy conservation error.

In such a situation the total energy in the chain is purely kinetic as the beads are drifting apart without touching each other. Therefore, we measure the kinetic energy at time $t=0$, before the the striking bead $0$ has any overlap with bead $1$, and after the chain has reached the steady state. This justifies neglecting the potential energy for calculating the energy conservation error.

\begin{figure}
\begin{center}
\includegraphics[width=0.35\textwidth]{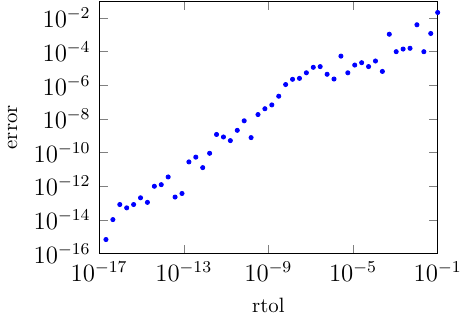}
\includegraphics[width=0.35\textwidth]{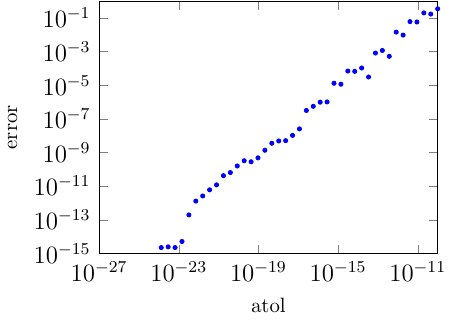}
\vspace{-0.2in}
   \caption{Energy conservation error as a function of the relative tolerance (left) and the absolute tolerance (right). For each plot the other tolerance was fixed at $0.0$.}
\label{fig:convergence}
\end{center}\end{figure}

In the VODE solver the estimated local error $\vec{e}$ is controlled by 
\begin{equation}
  \left| \frac{e_i}{\mathrm{EWT}_i} \right| \lesssim 1, \qquad i \in \{ 0, \dots, N-1 \} ,
\end{equation}
for each component of the error, where the error weights are defined as 
\begin{equation}
  \mathrm{EWT}_i = \mathrm{rtol}\times |z_i| + \mathrm{atol} ,
\end{equation}
where $\mathrm{rtol}$ and $\mathrm{atol}$ are the relative and absolute tolerances respectively \cite{BYR01}. Actually we would have to consider the two ghost particles here as well. However, we will neglect them for simplicity as their position and velocity variables do not change and therefore do not introduce errors.

One can set $\mathrm{rtol}=0$ or $\mathrm{atol}=0$ to have a purely absolute tolerance or purely relative tolerance respectively. Figure \ref{fig:convergence} shows the  energy conservation error as a function of the two tolerances. We observe that by setting strict tolerances in the simulation the error shrinks to nearly machine precision. This shows that our numerical scheme indeed conserves energy as would be expected from the physical system. In the simulations presented here we choose the tolerances such that the energy conservation error is $\mathcal{O}(10^{-4})$ or smaller. This has shown to be a good trade-off between accuracy and runtime.

\section{Stability under system perturbations}
The analysis of the dynamics of the beads in the vicinity of the interface shows that the numerically observed phenomena depend on the timing of gaps opening and closing and on the shape of the initially transmitted pulse. In experimental set-ups, the conditions might not be as well-controlled as in the numerical simulation due to the finite precision of the experimental apparatus. In order to investigate the stability of the previously observed effects under small perturbations of the chain parameters, the interaction potential exponent is randomly perturbed by
\begin{equation}
  n_i^{\mathrm{pert}} = n_i + n^\mathrm{rand}_i , \qquad i\in\{0, \dots, N-2\} ,
  \label{eqn:n_pert}
\end{equation}
where $n_i$ is the initial exponent and $n^\mathrm{rand}_i$ is taken from a Gaussian distribution with standard deviation $\sigma$ and mean $0$. However, the additional constraint $n_i^{\mathrm{pert}}>2.0$ is imposed in order to stay in the non-linear regime. The interaction exponent difference $\Delta n = n_2 - n_1$ provides a scale to which the introduced randomness can be compared. 

\begin{figure}
\begin{center}
\includegraphics[width=0.35\textwidth]{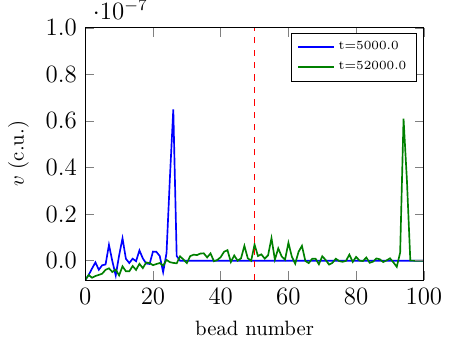}
    \includegraphics[width=0.35\textwidth]{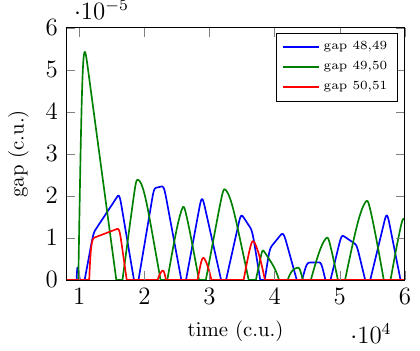}
    \vspace{-0.2in}
      \caption{Effect of perturbation \eqref{eqn:n_pert} for an interface with $\Delta n>0$, given $\sigma/\Delta n=0.05$.}
  \label{fig:stabilitysmalllarge}
\end{center}\end{figure}%

\begin{figure}
\begin{center}
    \includegraphics[width=0.35\textwidth]{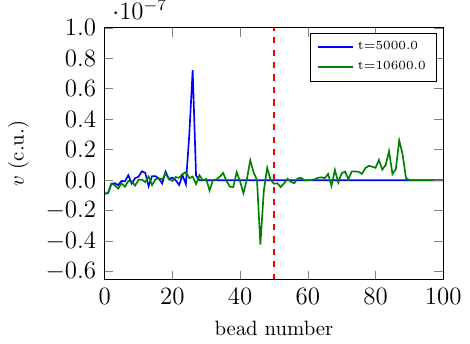}
    \includegraphics[width=0.35\textwidth]{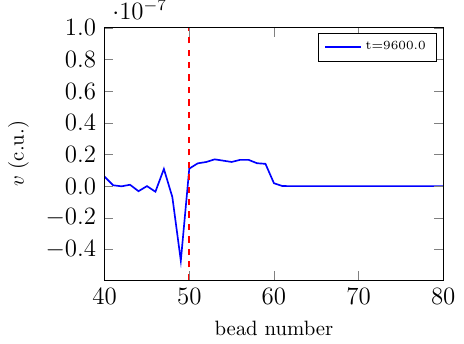}
    \vspace{-0.2in}
      \caption{Effect of perturbation \eqref{eqn:n_pert} with $\sigma/\Delta n=0.05$ on solitary wave propagation across an interface between media with $n_1=3.5$ and $n_2=3.0$: waves before and after hitting the interface (top) and collective excitation of the beads in the right medium (bottom). The interface is indicated by the dashed red line.}
  \label{fig:stabilitylargesmall}
\end{center}\end{figure}%
 
Figure \ref{fig:stabilitysmalllarge} shows the  effect of the perturbation for the propagation of a solitary wave across an interface with $\Delta n>0$. The first observation is that by introducing the perturbation, smaller waves behind the solitary wave develop, even in the monodisperse part of the chain. These smaller waves interfere with the two delayed secondary waves propagating away from the interface. However, a similar qualitative behaviour of the delayed secondary waves is observed up to the point where the smaller waves arising from the perturbations are of comparable size. Secondly, the gaps arising near the interface, which are responsible for the delayed transmitted and reflected secondary pulses, close faster due to interactions with waves arising from the impurities.

The effects of perturbation \eqref{eqn:n_pert} on the propagation across an interface with $\Delta n<0$ are shown in Figure \ref{fig:stabilitylargesmall}. We observe that the initial rectangular pulse, which causes the transmitted multipulse structure, develops for all tested perturbation strengths. Again, the agreement with the unperturbed case is determined by the relative sizes of smaller waves induced by the perturbations and the features arising from the interface, such that the leading peaks of the multipulse structure are clearly visible even for relatively large perturbations whereas the smaller tail is lost in the noise.

For $\Delta n>0$ the observed interface phenomena are present for perturbations of up to $1\%$ of the interaction potential difference $\Delta n$, whereas the multipulse structure and wave reflection for $\Delta n < 0$ are present for perturbations of more than $5\%$. Therefore, we anticipate that the effects at interfaces between different interaction potentials would be obtainable experimentally for both $\Delta n>0$ and $\Delta n<0$.


\begin{thebibliography}{10}
\expandafter\ifx\csname url\endcsname\relax
  \def\url#1{\texttt{#1}}\fi
\expandafter\ifx\csname urlprefix\endcsname\relax\def\urlprefix{URL }\fi
\expandafter\ifx\csname href\endcsname\relax
  \def\href#1#2{#2} \def\path#1{#1}\fi

\bibitem{JAE01}
H.~M. Jaeger, S.~R. Nagel, R.~P. Behringer, Granular solids, liquids, and
  gases, Rev. Mod. Phys. 68 (1996) 1259--1273.

\bibitem{SEN01}
S.~Sen, J.~Hong, J.~Bang, E.~Avalos, R.~Doney, Solitary waves in the granular
  chain, Physics Reports 462~(2) (2008) 21--66.

\bibitem{chong2017nonlinear}
C.~Chong, M.~A. Porter, P.~Kevrekidis, C.~Daraio, Nonlinear coherent structures
  in granular crystals, Journal of Physics: Condensed Matter 29~(41) (2017)
  413003.

\bibitem{rosas2018pulse}
A.~Rosas, K.~Lindenberg, Pulse propagation in granular chains, Physics Reports.

\bibitem{NES02}
V.~F. Nesterenko, Dynamics of heterogeneous materials, Springer, New York,
  2001.

\bibitem{JOH01}
K.~L. Johnson, Contact mechanics, Cambridge University Press, Cambridge, 1985.

\bibitem{job2005hertzian}
S.~Job, F.~Melo, A.~Sokolow, S.~Sen, How hertzian solitary waves interact with
  boundaries in a 1d granular medium, Physical review letters 94~(17) (2005)
  178002.

\bibitem{VER01}
L.~Vergara, Scattering of solitary waves from interfaces in granular media,
  Phys. Rev. Lett. 95 (2005) 108002.

\bibitem{VER02}
L.~Vergara, Delayed scattering of solitary waves from interfaces in a granular
  container, Phys. Rev. E 73 (2006) 066623.

\bibitem{KHA01}
D.~Khatri, D.~Ngo, C.~Daraio, Highly nonlinear solitary waves in chains of
  cylindrical particles, Granular Matter 14~(1) (2012) 63--69.

\bibitem{HER01}
H.~Hertz, Ueber die ber{\"u}hrung fester elastischer k{\"o}rper., Journal
  f{\"u}r die reine und angewandte Mathematik 92 (1882) 156--171.

\bibitem{LAN01}
L.~D. Landau, E.~M. Lif{\v s}ic, J.~B. Sykes, W.~H. Reid, Theory of elasticity,
  Pergamon Press, Oxford, 1963.

\bibitem{vorotnikov2017wave}
K.~Vorotnikov, Y.~Starosvetsky, G.~Theocharis, P.~Kevrekidis, Wave propagation
  in a strongly nonlinear locally resonant granular crystal, Physica D:
  Nonlinear Phenomena 365 (2017) 27--41.

\bibitem{COS01}
C.~Coste, E.~Falcon, S.~Fauve, Solitary waves in a chain of beads under hertz
  contact, Phys. Rev. E 56 (1997) 6104--6117.

\bibitem{NES03}
V.~F. Nesterenko, Propagation of nonlinear compression pulses in granular
  media, Journal of Applied Mechanics and Technical Physics 24~(5) (1983)
  733--743.

\bibitem{LAZ01}
A.~N. Lazaridi, V.~F. Nesterenko, Observation of a new type of solitary waves
  in a one-dimensional granular medium, Journal of Applied Mechanics and
  Technical Physics 26~(3) (1985) 405--408.

\bibitem{HAS01}
E.~Hasco{\"e}t, H.~J. Herrmann, Shocks in non-loaded bead chains with
  impurities, The European Physical Journal B - Condensed Matter and Complex
  Systems 14~(1) (2000) 183--190.

\bibitem{chatterjee1999asymptotic}
A.~Chatterjee, Asymptotic solution for solitary waves in a chain of elastic
  spheres, Physical Review E 59~(5) (1999) 5912.

\bibitem{hasan2017basic}
M.~A. Hasan, S.~Nemat-Nasser, Basic properties of solitary waves in granular
  crystals, Journal of the Mechanics and Physics of Solids 101 (2017) 1--9.

\bibitem{tang2017novel}
B.~Tang, Z.-H. Deng, K.~Deng, A novel envelope soliton solution to the granular
  crystal model, Communications in Theoretical Physics 68~(5) (2017) 627.

\bibitem{MAN01}
M.~Manciu, S.~Sen, A.~J. Hurd, Crossing of identical solitary waves in a chain
  of elastic beads, Phys. Rev. E 63 (2000) 016614.

\bibitem{FEL01}
F.~S. Manciu, S.~Sen, Secondary solitary wave formation in systems with
  generalized hertz interactions, Phys. Rev. E 66 (2002) 016616.

\bibitem{HON02}
J.~Hong, A.~Xu, Nondestructive identification of impurities in granular medium,
  Applied Physics Letters 81~(25) (2002) 4868--4870.

\bibitem{NES01}
V.~F. Nesterenko, C.~Daraio, E.~B. Herbold, S.~Jin, Anomalous wave reflection
  at the interface of two strongly nonlinear granular media, Phys. Rev. Lett.
  95 (2005) 158702.

\bibitem{POR01}
M.~A. Porter, C.~Daraio, I.~Szelengowicz, E.~B. Herbold, P.~Kevrekidis, Highly
  nonlinear solitary waves in heterogeneous periodic granular media, Physica D:
  Nonlinear Phenomena 238~(6) (2009) 666--676.

\bibitem{BUR01}
W.~P. Schonberg, H.~A. Burgoyne, J.~A. Newman, W.~C. Jackson, C.~Daraio,
  Proceedings of the 2015 hypervelocity impact symposium (hvis 2015) guided
  impact mitigation in 2d and 3d granular crystals, Procedia Engineering 103
  (2015) 52--59.

\bibitem{GAN01}
G.~Gantzounis, M.~Serra-Garcia, K.~Homma, J.~M. Mendoza, C.~Daraio, Granular
  metamaterials for vibration mitigation, Journal of Applied Physics 114~(9).

\bibitem{CHA01}
C.~S. Chang, J.~Gao, Non-linear dispersion of plane wave in granular media,
  International Journal of Non-Linear Mechanics 30~(2) (1995) 111--128.

\bibitem{LEO01}
A.~Leonard, C.~Daraio, Stress wave anisotropy in centered square highly
  nonlinear granular systems, Phys. Rev. Lett. 108 (2012) 214301.

\bibitem{MUE01}
N.~W. Mueggenburg, H.~M. Jaeger, S.~R. Nagel, Stress transmission through
  three-dimensional ordered granular arrays, Phys. Rev. E 66 (2002) 031304.

\bibitem{HON01}
J.~Hong, Universal power-law decay of the impulse energy in granular
  protectors, Phys. Rev. Lett. 94 (2005) 108001.

\bibitem{DON01}
R.~Doney, S.~Sen, Decorated, tapered, and highly nonlinear granular chain,
  Phys. Rev. Lett. 97 (2006) 155502.

\bibitem{MAN02}
M.~Manciu, Nonlinear acoustics of granular media, Ph.D. thesis, SUNY-Buffalo
  (2000).

\bibitem{DAR01}
C.~Daraio, V.~F. Nesterenko, E.~B. Herbold, S.~Jin, Energy trapping and shock
  disintegration in a composite granular medium, Phys. Rev. Lett. 96 (2006)
  058002.

\bibitem{POR02}
M.~A. Porter, C.~Daraio, E.~B. Herbold, I.~Szelengowicz, P.~G. Kevrekidis,
  Highly nonlinear solitary waves in periodic dimer granular chains, Phys. Rev.
  E 77 (2008) 015601.

\bibitem{BOE01}
N.~Boechler, G.~Theocharis, S.~Job, P.~G. Kevrekidis, M.~A. Porter, C.~Daraio,
  Discrete breathers in one-dimensional diatomic granular crystals, Phys. Rev.
  Lett. 104 (2010) 244302.

\bibitem{THE02}
G.~Theocharis, N.~Boechler, P.~G. Kevrekidis, S.~Job, M.~A. Porter, C.~Daraio,
  Intrinsic energy localization through discrete gap breathers in
  one-dimensional diatomic granular crystals, Phys. Rev. E 82 (2010) 056604.

\bibitem{HIN01}
E.~J. Hinch, S.~Saint{\textendash}Jean, The fragmentation of a line of balls by
  an impact, Proceedings of the Royal Society of London A: Mathematical,
  Physical and Engineering Sciences 455~(1989) (1999) 3201--3220.

\bibitem{ponson2010nonlinear}
L.~Ponson, N.~Boechler, Y.~M. Lai, M.~A. Porter, P.~Kevrekidis, C.~Daraio,
  Nonlinear waves in disordered diatomic granular chains, Physical Review E
  82~(2) (2010) 021301.

\bibitem{kim2018direct}
E.~Kim, A.~J. Mart{\'\i}nez, S.~E. Phenisee, P.~Kevrekidis, M.~A. Porter,
  J.~Yang, Direct measurement of superdiffusive energy transport in disordered
  granular chains, Nature communications 9~(1) (2018) 640.

\bibitem{br1}
G.~Theocharis, M.~Kavousanakis, P.~G. Kevrekidis, C.~Daraio, M.~A. Porter,
  I.~G. Kevrekidis, Localized breathing modes in granular crystals with
  defects, Physical Review E 80 (2009) 066601.

\bibitem{br2}
N.~Boechler, G.~Theocharis, S.~Job, P.~G. Kevrekidis, M.~A. Porter, C.~Daraio,
  Discrete breathers in one-dimensional diatomic granular crystals, Physical
  Review Letters 104 (2010) 244302.

\bibitem{br3}
G.~Theocharis, N.~Boechler, P.~G. Kevrekidis, S.~Job, M.~A. Porter, C.~Daraio,
  Intrinsic energy localization through discrete gap breathers in
  one-dimensional diatomic granular crystals, Physical Review E 82 (2010)
  056604.

\bibitem{LEN01}
D.~Leng, X.~Wang, G.~Liu, L.~Sun, Impulse absorption by horizontal magnetic
  granular chain, AIP Advances 6~(2).

\bibitem{MNJ01}
M.~Manjunath, A.~P. Awasthi, P.~H. Geubelle, Plane wave propagation in 2d and
  3d monodisperse periodic granular media, Granular Matter 16~(1) (2014)
  141--150.

\bibitem{TIC01}
A.~M. Tichler, L.~R. G\'omez, N.~Upadhyaya, X.~Campman, V.~F. Nesterenko,
  V.~Vitelli, Transmission and reflection of strongly nonlinear solitary waves
  at granular interfaces, Phys. Rev. Lett. 111 (2013) 048001.

\bibitem{leonard2014traveling}
A.~Leonard, C.~Chong, P.~G. Kevrekidis, C.~Daraio, Traveling waves in 2d
  hexagonal granular crystal lattices, Granular Matter 16~(4) (2014) 531--542.

\bibitem{2dhex}
T.~Hua, R.~A. Van~Gorder, Wave propagation and pattern formation in
  two-dimensional hexagonally-packed granular crystals under various
  configurations, arXiv preprint arXiv:1803.07190.

\bibitem{BYR01}
G.~D. Byrne, A.~C. Hindmarsh, A polyalgorithm for the numerical solution of
  ordinary differential equations, ACM Trans. Math. Softw. 1~(1) (1975) 71--96.

\end{thebibliography}
\end{document}